\documentclass[10pt,journal,compsoc]{IEEEtran}

\usepackage{amsmath}
\usepackage{subfigure}
\usepackage{graphicx}
\usepackage{textcomp}
\usepackage{xcolor}
\usepackage [english]{babel}
\usepackage [autostyle, english = american]{csquotes}
\MakeOuterQuote{"}
\usepackage{graphicx}
\usepackage{float}
\usepackage{bbm}
\usepackage{graphics} 
\usepackage{epsfig} 
\usepackage{tabularx}
\usepackage{multicol}
\usepackage{multirow}
\usepackage[noindentafter]{titlesec}
\usepackage{tabularx}
\usepackage{balance}
\usepackage{pdflscape}
\usepackage{bbding}
\usepackage{wasysym}
\usepackage{pifont}
\usepackage{afterpage}
\usepackage{capt-of}
\usepackage{rotating}
\usepackage{lscape}
\usepackage{hyperref}
\hypersetup{bookmarks=false}
\usepackage{pgfplots}
\usepackage{wrapfig}
\pgfplotsset{width=8cm,compat=1.9}
\graphicspath{ {images/} }
\usepackage{braket} 
\usepackage{algorithm}
\usepackage{xcolor}
\usepackage[noend]{algpseudocode}
\usepackage{mathtools,bm}
\usepackage{booktabs}

\DeclareMathOperator*{\argmax}{arg\,max}
\usepackage{enumitem}
\newcommand{\blue}[1]{\textcolor{black}{#1}}

\begin{document}
\urlstyle{tt}

\title{MCDS: AI Augmented Workflow Scheduling in Mobile Edge Cloud Computing Systems}

\author{
        Shreshth~Tuli,
        Giuliano~Casale
    and~Nicholas~R.~Jennings
\IEEEcompsocitemizethanks{
\IEEEcompsocthanksitem S. Tuli, G. Casale and N. R. Jennings are with the Department
of Computing, Imperial College London, United Kingdom.\protect
\IEEEcompsocthanksitem N. R. Jennings is also with Loughborough University, United Kingdom.\protect\\
E-mails: \{s.tuli20, g.casale\}@imperial.ac.uk, n.r.jennings@lboro.ac.uk.\protect
}
\thanks{Manuscript received ---; revised ---.}}

\markboth{IEEE Transactions on Parallel and Distributed Systems}%
{Tuli \MakeLowercase{\textit{et al.}}: --- }

\IEEEtitleabstractindextext{%
\begin{abstract}
Workflow scheduling is a long-studied problem in parallel and distributed computing (PDC), aiming to efficiently utilize compute resources to meet user's service requirements. Recently proposed scheduling methods leverage the low response times of edge computing platforms to optimize application Quality of Service (QoS).  However, scheduling workflow applications in mobile edge-cloud systems is challenging due to computational heterogeneity, changing latencies of mobile devices and the volatile nature of workload resource requirements. To overcome these difficulties, it is essential, but at the same time challenging, to develop a long-sighted optimization scheme that efficiently models the QoS objectives. In this work, we propose MCDS: Monte Carlo Learning using Deep Surrogate Models to efficiently schedule workflow applications in mobile edge-cloud computing systems. MCDS is an Artificial Intelligence (AI) based scheduling approach that uses a tree-based search strategy and a deep neural network-based surrogate model to estimate the long-term QoS impact of immediate actions for robust optimization of scheduling decisions. Experiments on physical and simulated edge-cloud testbeds show that MCDS can improve over the state-of-the-art methods in terms of energy consumption, response time, SLA violations and cost by at least 6.13, 4.56, 45.09 and 30.71 percent respectively.
\end{abstract}

\begin{IEEEkeywords}
AI for PDC, Edge computing, Cloud Computing, Deep Learning, Monte Carlo learning, Workflow scheduling.
\end{IEEEkeywords}}

\maketitle

\IEEEdisplaynontitleabstractindextext

\IEEEpeerreviewmaketitle

\IEEEraisesectionheading{\section{Introduction}\label{sec:introduction}}
\noindent
\IEEEPARstart{T}{he} explosive growth of data and the popularity of the Internet of Things (IoT) paradigm has given rise to new parallel and distributed computing (PDC) architectures like mobile edge-cloud computing~\cite{xu2018improved, gill2019transformative}. Edge-cloud computing, also referred to as fog computing, is an amalgamation of cloud and edge devices that aims to bring low-latency services to the users by leveraging nodes at the edge of the network and performing only the compute-intensive processing in the cloud~\cite{tuli2019fogbus}. The ability of edge computing to provide local compute, storage and network resources has increasingly been used to execute demanding and long-running workflow applications~\cite{impso, dnsga}. Unlike standalone workloads, workflows constitute several resource intensive and inter-dependent tasks, with their precedence constraints being modeled as a directed acyclic graph (DAG)~\cite{adhikari2019survey}. With the burgeoning shift to AI/ML/DL based workloads, modern workflow applications have become highly resource hungry~\cite{tuli2021cosco}.  Modern computing offers a new range of computationally heterogeneous workflows involving data gathering at the edge and processing at the cloud backend. Such workflows have recently been adapted to hybrid computing platforms like edge-cloud environments for their cost saving benefits~\cite{ding2018cost, matrouk2021scheduling}. The workflow scheduling problem aims to efficiently map the tasks onto the available resources to optimize the Quality of Service (QoS) metrics.

\textbf{Challenges.} A significant challenge faced in workflow scheduling is the requirement of handling heterogeneous and non-stationary user demands~\cite{machen2016migrating, ben2020edge}. As edge and cloud nodes have different computational capacities, network bandwidth and latencies, workflow schedulers need to be aware of the resource heterogeneity to make optimal decisions. Further, the Service Level Agreement (SLA) deadlines of QoS objectives may be different for distinct applications or may even vary across the input processing requests for the same application. For instance, many industries, especially healthcare, robotics and smart-cities demand ultra-low response times for SLA sensitive tasks~\cite{haussmann2019cost}. Battery driven mobile edge computing nodes also have strict energy consumption demands~\cite{choi2019optimizing, tuli2021generative}. Applications deployed on a pay-per-use cloud service model aim at reducing the overall execution costs. Further, the problem of workflow scheduling is exacerbated by the dynamism of the system and the inconsistent DAG dependencies across different long-running workflows, making long-term optimization hard. 

The dynamism in the system arises from two primary sources: task volatility and host mobility. Task volatility covers the changing resource requirements or service demands of tasks. Host mobility entails the changing geographic location of host machines that can give rise to non-stationary latencies.  With the rising use of mobile computational devices, the network characteristics like latency and bandwidth have become volatile. This requires real-time monitoring and updating the scheduling decisions, also known as rescheduling in literature~\cite{yu2005cost, tuli2021pregan}. As scheduling decisions change for active tasks, rescheduling might entail live task migration, also requiring schedulers to consider migration overheads. To avoid excessive overheads, the objective score estimation of such methods needs to be extremely accurate~\cite{tuli2021cosco}.

\textbf{Existing solutions.} Over the past few years, several workflow schedulers have been proposed that aim to tackle the above mentioned challenges~\cite{dnsga, impso, esvr, closure}. Many of these methods use Artificial Intelligence (AI) based optimization schemes like evolutionary schemes or reinforcement learning~\cite{alkhanak2015cost}. Some classical methods also use heuristics or meta-heuristics to optimize their decisions~\cite{genez2012workflow, xu2009multiple}. However, most of these methods solve the challenges of workflow scheduling to a limited extent. Traditional reinforcement learning methods like Q-Learning are typically brittle in terms of the environment assumptions that makes them unsuitable for highly volatile scenarios~\cite{haussmann2019cost}. Other methods like genetic algorithms have been shown to have the agility and adaptability required for dynamic workflow demands~\cite{dnsga}. However, non-local jumps in many evolutionary algorithms lead to several task migrations, increasing network overheads and degrading their performance~\cite{tuli2021cosco}. Other swarm optimization based methods that divide the timeline into periodic scheduling intervals use neural networks to approximate QoS scores of the next interval~\cite{impso, saeedi2020improved}. However, such methods are myopic as they optimize short-term objective scores, leading them to often get stuck at a local optimum~\cite[Chapter~7]{sutton2018reinforcement}. Further, such methods often struggle to produce accurate QoS estimates in dynamic settings (with volatile tasks and mobile hosts) to execute apt decision optimization~\cite{song2020win}.

\textbf{Background and new insights.} To mitigate the risks of short-term goal optimization and generate a myopic QoS estimate, we develop on some recent ideas. In the past, schedulers have used deep neural networks to approximate the optimization objectives~\cite{dnsga, tuli2021cosco}. Such neural networks are called deep surrogate models (DSMs) as they act as approximators for the QoS scores that are otherwise time-consuming to obtain (requiring physical execution of the scheduling decisions)~\cite{tuli2021cosco}. Deep surrogate models provide an improved QoS due to their ability to accurately map objective scores and quickly adapt to diverse scenarios. However, due to the poor state-space exploration, such methods can perform poorly in settings unseen during model training~\cite{tuli2021cosco}. 
As surrogate models only provide a single-step objective estimate, optimizing using such models can adversely affect long-term QoS scores. Thus, some reinforcement learning methods use look-ahead search schemes to get a long-term score estimate~\cite{sutton2018reinforcement}. One such method is Monte Carlo learning that runs several multi-step executions to get an unbiased estimate of the long-term effect of an immediate action~\cite{sutton2018reinforcement}. Running such executions in a real environment is infeasible due to the large execution times. As scheduling decisions need to be taken in near real-time, prior work does not use such look-ahead schemes and hence performs myopic optimization. However, recently proposed co-simulation methods seem like a promising solution for this problem~\cite{tuli2021cosco}. Co-simulation allows schedulers to leverage event-driven simulators to quickly get a QoS estimate instead of executing decisions on a physical platform. 

\textbf{Our contributions.} Combining Monte Carlo learning and deep surrogate models for QoS optimal workflow scheduling is non-trivial. This is because vanilla Monte Carlo learning requires the computationally expensive execution of several multi-step simulation runs with random scheduling actions. This may be infeasible when frequent decision making is required, especially for volatile environments~\cite{haussmann2019cost}. Further, traditional deep surrogate models trained on historical data only provide the QoS score of the system state immediately after executing an action and not a long-term QoS. To address these issues, instead of randomly choosing the actions to simulate, we train a deep surrogate model to approximate long-term QoS scores within a Monte Carlo learning framework. To discover potentially better actions, we constantly explore the state-action search space. To start with, we use the action output of a short-term optimization scheme. 

Drawing all these insights together, we present MCDS: Monte Carlo Learning using Deep Surrogate models for workflow scheduling. MCDS is an AI augmented scheduler that uses Upper-Confidence-Bound (UCB) exploration~\cite[Chapter~2]{sutton2018reinforcement} and short-term optimization action output as domain knowledge. Running co-simulations on several multi-step decisions allows MCDS to get a long-term QoS estimate of each immediate action choice. Once we have a trained deep surrogate model, we employ a gradient optimization using the back-propagation to input (GOBI) scheme to reach optimal decisions quickly~\cite{tuli2021cosco}.  To test the efficacy of the MCDS policy, we perform experiments on a physical mobile edge-cloud setup with ten hosts and a simulated setup with fifty hosts with real-world application workloads. The set of schedulers used in our experiments includes MCDS and four other state-of-the-art baselines. Our evaluations show that MCDS performs \textit{best} in terms of QoS parameters, giving at least 6.13\%, 4.56\%, 45.09\% and 30.71\% better energy consumption, response time, SLA violation rate and execution cost respectively. These results are obtained with up to 13.19\% lower scheduling time than the baseline with the second-best QoS scores.

The rest of the paper is organized as follows. Section~\ref{sec:related_work} overviews the related work. Section~\ref{sec:system} outlines the system model assumptions. Section~\ref{sec:surrogate} details the deep surrogate model in the MCDS scheduler. Section~\ref{sec:montecarlo} presents the Monte Carlo learning to train and generate scheduling decisions. A performance evaluation of the proposed method is shown in Section~\ref{sec:perf_eval}. Finally, Section~\ref{sec:conclusions} concludes.

\section{Related Work}
\label{sec:related_work}

We now analyze the prior work in more detail, dividing them into two classes: heuristic and meta-heuristic approaches. 

\textbf{Heuristic methods.} Several classical works use heuristics to optimize the workflow scheduling decisions~\cite{byun2011cost, chen2017efficient, chirkin2017execution}. Many methods in this category use heuristics motivated by the objectives they aim to optimize, such as average response time, energy consumption, execution cost, resource utilization or SLA violation rates. Some methods partition workflows into multiple sub-workflows to minimize expected communication overheads that facilitate the reduction of overall response time~\cite{chen2017efficient}. Other methods approximate objectives like energy consumption by easily measurable metrics like CPU utilization to output a scheduling decision~\cite{alkhanak2015cost}. Some approaches formulate the workflow scheduling task as a simplified problem like subset-sum and use mean clustering techniques~\cite{singh2018novel}.  However, all of these methods fail to perform well in highly volatile settings and only aim at optimizing specific objectives (via heuristics) that make them unsuitable for demanding modern-day applications~\cite{adhikari2019survey}.

\textbf{Meta-Heuristic methods.} This class of methods leverages high-level problem independent algorithms to find the optimal scheduling decision for the workflows. Most state-of-the-art approaches belong to this category. Among these, many use variants of the Particle Swarm Optimization (PSO) technique~\cite{pandey2010particle, impso}. Such a technique starts with several random or heuristically initialized candidate solutions. Each candidate is iteratively optimized, moving it slightly in the state-space where the optimization objective tends to increase. One of the baselines in our experiments is the immune based particle swarm optimization (IMPSO) method~\cite{impso}. This is because IMPSO uses candidate affinity to prevent poor candidates from being discarded in subsequent iterations, allowing it to surpass other PSO based methods in terms of execution costs and average response time. Other techniques, categorized commonly as list scheduling, use metrics like earliest finish time, critical path, and dynamic resource utilization levels~\cite{topcuoglu2002performance, adhikari2019survey}. However, list scheduling performs poorly in settings with non-preemptable jobs and heterogeneous requirements or machines~\cite{adhikari2019survey}.

Other recent methods use genetic algorithms to optimize the scheduling decision. Most such works use a neural network to get an estimate of the QoS objectives~\cite{dnsga, esvr}. As in swarm based methods, the algorithm uses several candidate solutions. However, these algorithms can lead to non-local jumps in the search space due to operations like cross-over and mutations. This can lead to better QoS estimates, but also tends to have higher task migration overheads as converged decisions can be very different from the decision of the previous scheduling interval~\cite{tuli2021cosco}. For our experiments, we use ESVR~\cite{esvr} and DNSGA~\cite{dnsga} as our baselines from this class. ESVR initializes its candidate population using the Heterogeneous Earliest Finish Time (HEFT) heuristic and optimizes using the crossover-mutation scheme. To account for volatility in the system, ESVR continuously fine-tunes the neural network surrogate using the latest workload traces and host characteristics~\cite{esvr}. DNSGA is a multi-objective optimization method that uses a Pareto Optimal Front (POF) aware approach that prevents the set of candidates from converging to the same optima~\cite{dnsga}. Ismayilov et al. and Pham et al. show that these two methods out-perform previously proposed genetic algorithms based techniques and hence are used as baselines in our experiments~\cite{dnsga, esvr}.

Finally, another recently proposed workflow scheduling model, namely Closure, uses an attack-defense game-theoretic formulation~\cite{closure}. Unlike other schemes that assume mostly homogeneous resources, Closure has been shown to efficiently manage heterogeneous devices by calculating the Nash Equilibrium of the attack-defense game model. This is crucial in edge-cloud environments where there are contrasting resource capacities of edge and cloud nodes. This method is also one of the baselines in our experiments.

\section{System Model and Problem Formulation}
\label{sec:system}

\begin{figure}
    \centering  \setlength{\belowcaptionskip}{-20pt}
    \includegraphics[width=0.75\linewidth]{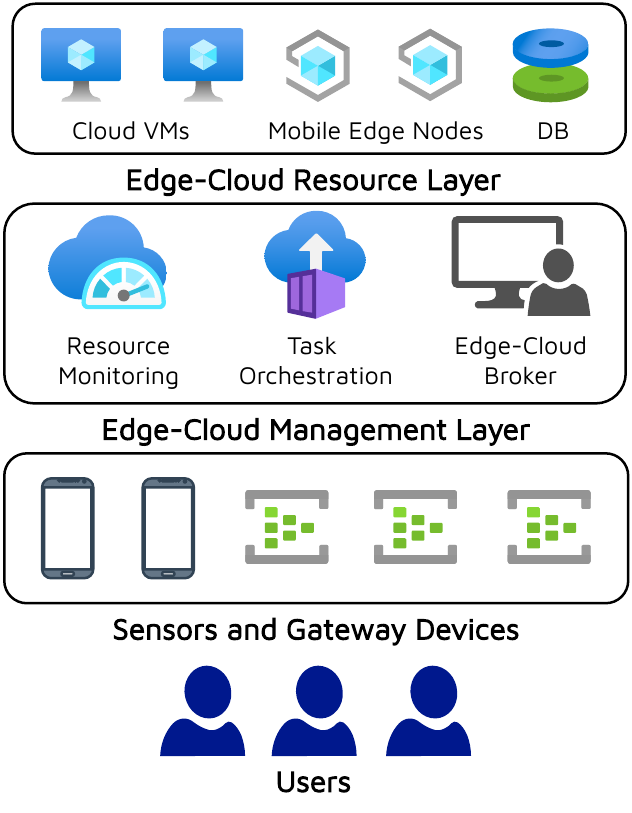}
    \caption{System Model.}
    \label{fig:system} \vspace{-10pt}
\end{figure}

\begin{figure*}[!t]
    \centering \setlength{\belowcaptionskip}{-20pt}
    \includegraphics[width=\linewidth]{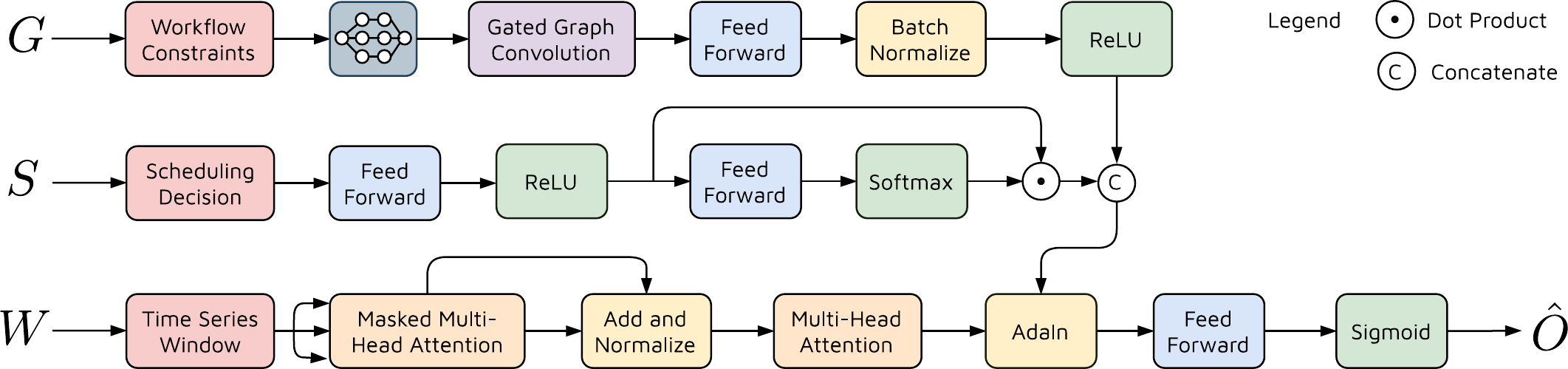}
    \caption{Deep Surrogate Model used in MCDS Scheduler. The three inputs to the model are shown in red. Feed-forward, graph convolution and attention operations are shown in blue, purple and orange. All activations are shown in green and all non-parameterized layers in yellow.} \vspace{-10pt}
    \label{fig:model}
\end{figure*}

\subsection{Environment Model}

In this work, we assume a typical mobile edge-cloud computing setup with multiple heterogeneous edge or cloud nodes in a master-slave fashion~\cite{tuli2019fogbus, ahuja2012survey}. We give an overview of the MCDS system model in Figure~\ref{fig:system}. The resource layer in our system includes edge nodes in the same local area network as the edge-cloud broker and cloud machines at a multi-hop distance from the broker. The edge nodes are assumed to be mobile, whereas the nodes at the cloud are assumed to be fixed in terms of their geographical location. All data inputs and outputs are asynchronously shared over a common database (DB). We consider workflows that include a batch of tasks and their corresponding precedence constraints as a DAG. In a physical deployment, these batches are passed on from sensors and actuators to the broker using gateway devices. The broker resides in the management layer and takes all the scheduling and migration decisions for each new and active task. We also assume that the resource capacities like that of the CPU, RAM, Bandwidth and Disk of each node are known in advance, and similarly, we can sample the resource consumption for each task in the environment at any time. We periodically measure the utilization of CPU, RAM, Bandwidth and Disk for each task and the corresponding workflow in the system. Further, for each task, a soft service level deadline is defined at the time the workflow is sent to the edge-cloud environment.  All workflows are received from an IoT layer that includes sensors and actuators to collect data from the users and send it to the broker via the gateway devices. The broker then decides where to allocate new tasks and where to migrate active tasks in the system, if required.

\subsection{Workload Model}

We assume scheduling as a discrete-time control problem as standard in prior work~\cite{tuli2021cosco, basu2019learn, tuli2021start}. We divide the timeline into equal duration intervals, with the $t$-th interval denoted as $I_t$ (starting from $i=0$). We assume that there are a fixed number of host machines, the set of which is denoted as $H$. Each workflow is modeled as a directed acyclic graph $G = (V, E)$, where $V$ denotes the tasks as nodes and $E$ denotes the execution precedence constraints as directed edges. Each edge $(i,j)$ in $E$ is an ordered pair such that task $i$ must complete execution before task $j$ is scheduled.  We also consider that new workflows created at the interval $I_t$ are denoted as $N_t$, with all active workflows being denoted as $A_t$. A workflow is considered to be active if at least one task of that workflow is being executed in the edge-cloud environment. If no task of a workflow $G \in N_t$ can be allocated to an edge or cloud node then it is added to a wait queue $W_t$. All created workflows that are not active and are not in the wait queue are considered to be completed workflows and we can calculate their QoS metrics like response time and SLA violation.

\subsection{Problem Formulation}

We denote the QoS objective score for interval $I_t$ as $O_t$ that we need to maximize. We denote the utilization metrics of all hosts in the interval $I_{t-1}$ as $U_t$. We collect the DAGs of all active workflows $A_{t-1}$ and denote them as $G_t$ (to represent the workflow constraint graph). Now using $U_t$ and $G_t$, we need to predict a scheduling decision $S_t$. All tasks with no incomplete dependent tasks in $N_t \cup W_t \cup A_t$ are called feasible tasks. $G_t$ includes allocation decisions for feasible and migration decisions for active tasks in the system. Thus the problem can be formulated as:
\begin{equation}
\label{eq:problem}
\begin{aligned}
& \underset{S_t}{\text{maximize}}
& & \sum_{t=0}^{\infty} O_t \\
& \text{subject to}
& & \forall\ t, S_t : P_t \cup Q_t \rightarrow H, \\
&&& \forall\ t, P_t = \text{ set of feasible tasks in }N_t \cup W_t \cup A_t, \\
&&& \forall\ t, Q_t = \text{ set of active tasks in the system}.
\end{aligned}
\end{equation}
\section{Deep Surrogate Model}
\label{sec:surrogate}
As discussed in Section~\ref{sec:introduction}, our MCDS scheduler uses a deep neural network as a surrogate model. The MCDS scheduler lies in the edge-cloud broker (Figure~\ref{fig:system}). We first describe the working of this model and how we use it to obtain an estimate of the objective score. To do this, we encapsulate the state of the system using the host characteristics $U_t$ and the workload constraints $G_t$. To predict an estimate of the objective score, we provide the model with the system state and the scheduling decision. Instead of using $U_t$, we use a time-series sliding window of size $k$, denoted as $W_t$, to capture temporal trends of the host utilization characteristics. Thus, at the start of the interval $I_t$, the input to the surrogate model is $[G_t, S_t, W_t]$. Without loss in generality, we drop the subscript $t$ for the rest of the discussion. An overview of the deep neural network used as a surrogate model in the MCDS scheduler is shown in Figure~\ref{fig:model}.  The constituent model components are detailed next.

\subsection{Workflow Constraint Graph Encoding}
To generate a QoS estimate, we encode the workflow constraint graph $G$ using a gated-graph convolution network (GGCN). This is because GGCN has been shown to be effective in capturing dynamically changing graph like inputs~\cite{li2015gated}. We first form an embedding for each node in the graph; for active tasks we use the SLA deadline and the resource utilization characteristic and for inactive tasks we use the zero-vector. We denote this embedding for a task $i$ as $e_i$. Following \blue{\cite{tuli2021hunter}}, the gating stage is realized as a Gated Recurrent Unit (GRU) resulting in \emph{graph-to-graph} updates as: 
\begin{align}
\begin{split}
    r_i^{0} &= \mathrm{Tanh} \left( \blue{w}\ e_{i} + b \right),\\
    x^k_i &= \sum_{j \in n(i)} \blue{w}^k r_{j}^{k-1} ,\\
    r^k_{i} &= \mathrm{GRU} \left( r^{k-1}_i, x^{k}_{i} \right).\\
\end{split}
\end{align}
\blue{Here, $w$ and $b$ are the weight matrix and bias parameter of the neural network and $k$ is the iteration index that varies from 1 to $p$, each corresponding to a graph convolution operation.} Further, the messages for task $i$ are aggregated over one-step connected neighbors $n(i)$ over $p$ convolutions, resulting in an embedding $r^{p}_{i}$ for each task node in the graph. The stacked representation for all tasks is represented as $r^p$ \blue{and is used in the downstream model}. We generate the workflow graph encoding $e^G$ by passing $r^p$ through a feed-forward layer with $\mathrm{ReLU}$ activation as
\begin{equation}
    e^G = \mathrm{ReLU}(\mathrm{BatchNorm}(\mathrm{FeedForward}(r^p))),
\end{equation}
where $\mathrm{BatchNorm}$ standardizes the output of the $\mathrm{FeedForward}$ layer for the input batch.

\subsection{Scheduling Decision Encoding}
We also encode the scheduling decision for $p$ tasks, first as a matrix $M^S$ and then pass it through a feed-forward layer. Then we apply \textit{Bahdanau style} self-attention~\cite{bahdanau2015neural} to reduce its dimensions size and generate the scheduling decision encoding $e^S$. Thus, 
\begin{align}
\begin{split}
    M &= \mathrm{ReLU}(\mathrm{FeedForward}(M^S)),\\
    attn &= \mathrm{Softmax}(\mathrm{FeedForward}(M)),\\
    e^S &= \sum_{i=1}^p attn_i \cdot M_i.
\end{split}
\end{align}
Here, $M_i$ is the one-hot vector corresponding to the $i$-th tasks out of $p$ tasks in the system. 
This operation takes a convex combination of the multiple scheduling decisions to create a single encoding that encapsulates the inter-task dependencies. This relieves the downstream predictors from discriminating among scheduling decisions. 

\subsection{Time Series Window Encoding}
For any three input matrices $Q$, $K$ and $V$, we define Multi-Head Self Attention~\cite{vaswani2017attention} as passing it through $h$ (number of heads) feed-forward layers to get $Q_i$, $K_i$ and $V_i$ for $i \in \{1, \ldots, h\}$, and then applying attention as
\begin{align}
\begin{split}
    \mathrm{MultiHeadAtt}(Q, K, V) &= \mathrm{Concat}(H_1, \ldots, H_h),\\
    \text{where } H_i &= \mathrm{Attention}(Q_i, K_i, V_i),
\end{split}
\end{align}
where $\mathrm{Attention}(Q_i, K_i, V_i)$ is the scaled-dot product attention operation~\cite{vaswani2017attention}. For a time-series input of host resource utilization characteristics, we perform transformer like multi-head attention operations as:
\begin{align}
\begin{split}
    W^1 &= \mathrm{Mask}(\mathrm{MultiHeadAtt}(W, W, W)), \\
    W^2 &= \mathrm{LayerNorm}(W + W^1),\\
    e^W &= W^2 + \mathrm{MultiHeadAtt}(W^2, W^2, W^2).
\end{split}
\end{align}
The first attention operation is masked to prevent the encoder from looking at the datapoints for future timestamp values at the time of training as all time-series windows are given at once to allow parallel training. $\mathrm{LayerNorm}$, unlike $\mathrm{BatchNorm}$, performs normalization across the values in the input instead of the batch.  Multi-head attention allows the model to make the subsequent prediction computationally simpler, enabling higher model scalability and generalizability~\cite{lee2019attention}. The final window encoding obtained is $e^W$.

\subsection{Estimating the Objective Score}
Now that we have encoded representations of the host characteristics ($e^W$), scheduling decision ($e^S$) and the workflow tasks ($e^G$), we use Adaptive Instance Normalization (AdaIn)~\cite{huang2017arbitrary} to normalize our window encoding as:
\begin{align}
\begin{split}
    e^{GS} &= [e^G; e^S], \\
    e &= \mathrm{AdaIn}(e^W, e^{GS}).\\
\end{split}
\end{align}
Here, $e^{GS}$ is the concatenation of $e^G$ and $e^S$. \blue{Motivated from style-transfer neural networks, AdaIn adaptively normalizes the time-series encoding  using the workflow and scheduling decision encodings.} This allows the model to maintain robustness in scenarios with highly time-varying host characteristics. The final encoding is then passed through a feed-forward layer to generate the objective score estimate
\begin{equation}
    \hat{O} = \mathrm{Sigmoid}(\mathrm{FeedForward}(e)).
\end{equation}
Unlike traditional deep surrogate model based schedulers, the output of our model is an estimate of the long-term objective score instead of the short-term (single scheduling interval) score. This allows the scheduler to avoid local optima when making sequential scheduling decisions~\cite{sutton2018reinforcement}.  We denote this surrogate model as a function $f(G, W, S) = \hat{O}$. We use a Monte Carlo approach to train this model as described next.

\begin{figure*}
    \centering \setlength{\belowcaptionskip}{-12pt}
    \includegraphics[width=\linewidth]{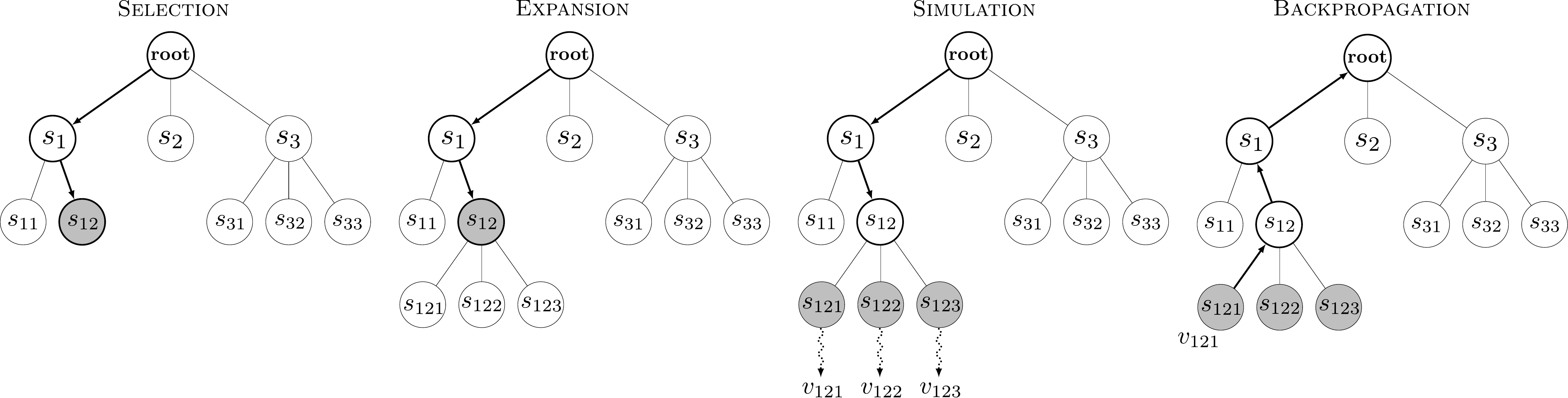}
    \caption{Building the Monte Carlo Tree. The selection operation increments the $n$ values of the nodes. The expansion operation sets the $q$ values using the immediate QoS score for each action. The backpropagation operation updates the $v$ values of the nodes up to the root.}
    \label{fig:mcts}
\end{figure*}

\section{Monte Carlo Learning}
\label{sec:montecarlo}

We develop a modified version of the Monte Carlo Tree Search (MCTS) to train our deep surrogate model and also actively take scheduling decisions. MCTS is a method of building a search tree considering the current state as the root node~\cite{silver2009monte}. The nodes in this tree denote states of the system and edges denote actions. Each node is assumed to be either a leaf node or have one or more child nodes corresponding to the feasible actions from this node\footnote{In our discussion, we use action and scheduling decision interchangeably.}. For our Monte Carlo tree, we consider that each node is defined by the following set of variables: $[G, W, S, q, v, n]$. Here, $G$, $W$ and $S$ represent the workflow constraints, time-series window and last scheduling decision; $q, v, n$ denote the short-term reward, \textit{value function} and the number of visits to the node. \blue{The value function is the long-term estimate of the QoS for the node.} Building this tree involves finding the feasible actions from a state, executing several simulation steps and iteratively updating the root node with the best action. 

\subsection{Building the Monte Carlo Tree}
\label{sec:build}

The process of building the Monte Carlo tree involves iteratively performing four operations: Selection, Expansion, Simulation and Backpropagation. We start with only the root node (initial state), with $q, v$ initialized as 0 and $n$ initialized as 1. An overview of these operations is presented in Figure~\ref{fig:mcts}.

\textbf{Selection.} This operation selects a leaf-node of the tree. Starting at the root node, we recursively select the child node that has the highest Upper-Confidence-Bound (UCB). For the current node $[G, W, S, q, v, n]$, we select one of the child nodes $\{[G_i, W_i, S_i, q_i, v_i, n_i]\}_i$ using the rule~\cite{aima}
\begin{equation}
\label{eq:selection}
    \text{selected node} = \argmax_i v_i + \sqrt{\frac{c_1 \times \ln{n}}{n_i}}.
\end{equation}
Here the first term corresponds to the exploitation part of the search algorithm and the second term corresponds to exploration. $c_1 \in [0, 1]$ is the exploration parameter. The $n$ value of each node that is selected in this recursive process is incremented by 1. When only the root node is present in the tree, we return that as the selected node. 

\textbf{Expansion.} After the selection process, we reach a leaf node of the tree. We use a co-simulator to find out all possible actions $\{S_i\}_i$ from this node and expand it to generate a new node for each such decision. The co-simulator also provides us the $G$ and $W$ for the new nodes. For each new node in the tree, $v, n$ are initialized as 0, 1, respectively. The $q$ value is set as the QoS objective score returned by the co-simulator for this decision in one scheduling interval.

\textbf{Simulation.} For the new nodes in our tree, we use the co-simulator to execute a sequence of $\phi$ actions to get a long-term QoS estimate of this node, where $\phi$ is called the roll-out parameter. This QoS value is set as the $v$ value of this node. While training, the actions used in these simulations are based on a random scheduler. At test time, we use the deep surrogate model based scheduler. More details on this in Sections~\ref{sec:training} and~\ref{sec:testing}. 

\textbf{Backpropagation.} Now that we have the updated $v$ values of the new leaf nodes, we backpropagate these up to the root for updating the value functions. For every node ($[G, W, S, q, v, n]$) other than the leaf-nodes in the path to the root node with child nodes $\{[G_i, W_i, S_i, q_i, v_i, n_i]\}_i$, we propose the weighted update rule
\begin{equation}
\label{eq:backprop}
    v = q + \frac{\sum_i n_i \cdot v_i}{\sum_i n_i},
\end{equation}
\textit{i.e.}, the sum of $q$ and the weighted average of the $v$ values of child nodes. The $q$ value here gives us the immediate reward and weighted-average gives us the estimate of the long term reward once the action corresponding to this node is taken~\cite{aima}.

We perform the four operations: selection, expansion, simulation and backpropagation a fixed number of times ($\psi$) to generate the Monte Carlo Tree. The expansion operation adds possibly one or more nodes to the tree in each iteration. The UCB metric \blue{in the selection phase} trades off between exploration of new action choices and exploitation of value function estimates in each iteration. As $\psi$ increases, the value function estimates become closer to the expected empirical objective scores.

\begin{figure}[t]
    \centering  \setlength{\belowcaptionskip}{-12pt}
    \includegraphics[width=\linewidth]{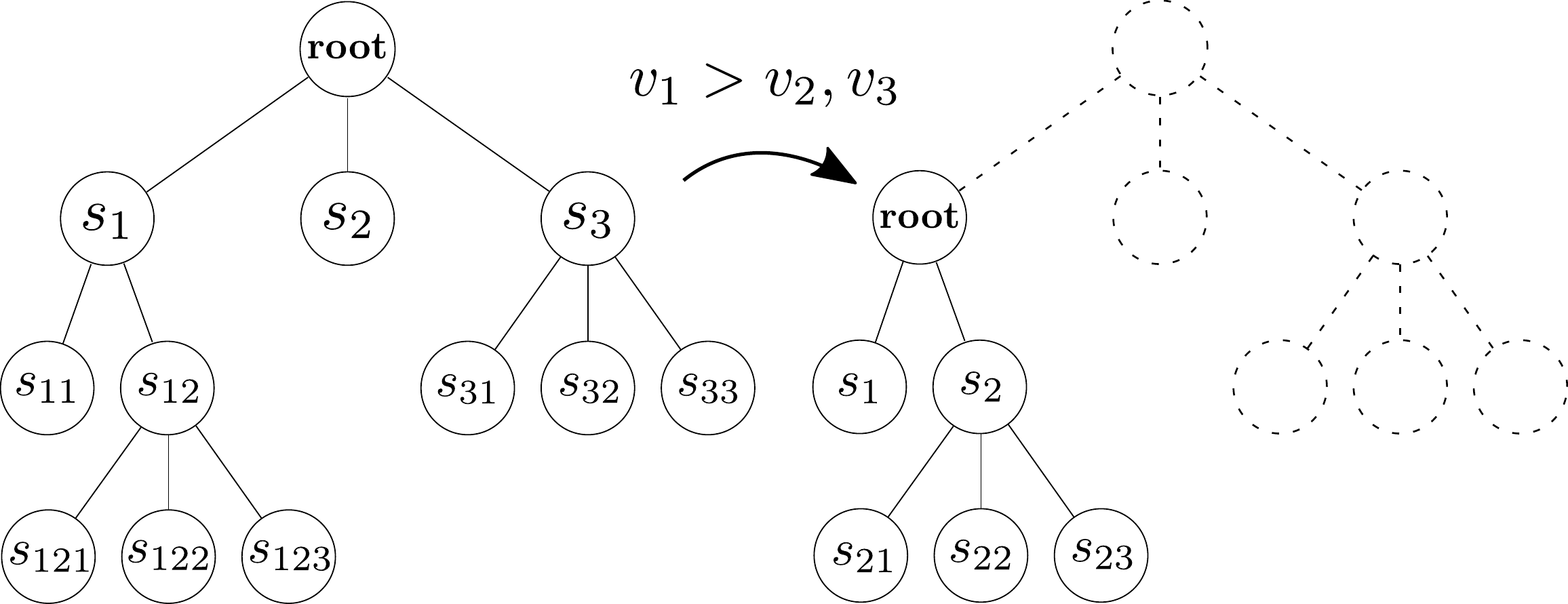}
    \caption{Choosing best action and updating the Monte Carlo Tree after simulations.}
    \label{fig:mcts_update}
\end{figure}

\begin{algorithm}[t]
    \begin{algorithmic}[1]
    \Require
    \Statex Number of simulation steps $\psi$; Rollout parameter $\phi$; Dataset used for training $\Lambda$; Deep Surrogate Model $f$
    \State Initialize $root$ node
    \Procedure{MCTS-Train}{$root$}
        \State \textbf{for} $i \gets 0$ to $\psi$ \textbf{do}
        \State \hspace{\algorithmicindent} Recursively select child nodes using Eq.~\eqref{eq:selection} \label{line:selection}
        \State \hspace{\algorithmicindent} Expand leaf-node $\forall$ possible actions  
        \State \hspace{\algorithmicindent} Using random policy, run $\phi$-step co-simulations 
        \Statex \hspace{4em} to generate QoS scores
        \State \hspace{\algorithmicindent} Recursively back-propagate scores using Eq.~\eqref{eq:backprop}        
    \EndProcedure
    \Procedure{Update}{$root$}
        \State $[G, W, S, q, v, n] \gets root$
        \State $\{[G_j, W_j, S_j, q_j, v_j, n_j]\}_j \gets$ child nodes of $root$
        \State $i = \argmax_j v_j$ \label{line:argmax}
        \State Add $(f(G, W, S), O_t + v_i)$ to $\Lambda$ \label{line:data}
        \State $root \gets [G_i, W_i, S_i, q_i, v_i, n_i]$
        \State \textbf{return} $S_i$
    \EndProcedure
    \Procedure{Train}{scheduling interval $I_t$}
        \State $\textsc{MCTS-Train}(root)$
        \State $S_i \gets \textsc{Update}(root)$
        \State Tune $f$ using the dataset $\Lambda$ using~\eqref{eq:loss} as loss
        \State \textbf{return} $S_i$
    \EndProcedure
    \end{algorithmic}
\caption{Training the deep surrogate model}
\label{alg:training}
\end{algorithm}

\subsection{Model Training}
\label{sec:training}

To train the deep surrogate model, at every interval $I_t$ we build the tree as described in Section~\ref{sec:build} as shown by the $\textsc{MCTS-Train}$ function in Algorithm~\ref{alg:training}. (we start with a single node in $I_0$). We use a random scheduler to execute the \textit{simulation} operation. The random scheduler gives us an unbiased estimate of the long-term value function of a leaf-node (line~\ref{line:selection} in Alg.~\ref{alg:training}). After performing the operations $\psi$ number of times, we choose the action $S_i$ corresponding to child node of the root with the highest value-function $v_i$ (line~\ref{line:argmax} in Alg.~\ref{alg:training}). For training, instead of using the $q_i$ value of this node, we get the QoS score in the next interval $I_{t+1}$ \textit{i.e.} $O_t$ and add the datapoint $(f(G, W, S), O_t + v_i)$ to the dataset (line~\ref{line:data} in Alg.~\ref{alg:training}). For the next interval, we update the Monte Carlo tree by changing the root node to the selected child node (see Figure~\ref{fig:mcts_update} and $\textsc{Update}$ function in Alg.~\ref{alg:training}). This allows us to re-use co-simulation data in the subsequent scheduling intervals. 

Using the collected dataset, we use experience-replay~\cite{rolnick2019experience} to decouple training from the execution of the workloads on our edge-cloud setup. We update the weights of the deep surrogate model using the loss:
\begin{equation}
\label{eq:loss}
    \mathrm{Loss} = \mathrm{MSE}(f(G, W, S), O_t + v_i),
\end{equation}
where $\mathrm{MSE}$ denotes mean-square-error and $G, W$ are the workflow constraints and time-series host characteristics of the root node state. We use a cross-validation style training process with early-stopping as the convergence criterion.

\subsection{Scheduling using the Trained Model}
\label{sec:testing} 
\begin{algorithm}[t] 
    \begin{algorithmic}[1]
    \Require 
    \Statex Number of simulation steps $\psi$; Learning rate $\gamma$; 
    \Statex Convergence threshold $\epsilon$; Iteration limit $L$
    \Statex Pre-trained Deep Surrogate Model $f$
    \State Initialize $root$ node
    \Procedure{MCTS-Test}{$root$}
        \State \textbf{for} $i \gets 0$ to $\psi$ \textbf{do}
        \State \hspace{\algorithmicindent} Recursively select child nodes using Eq.~\eqref{eq:selection_test} 
        \State \hspace{\algorithmicindent} Expand leaf-node $\forall$ possible actions  
        \State \hspace{\algorithmicindent} Use $f(G, W, S)$ as generate QoS scores \label{line:simulation2}
        \State \hspace{\algorithmicindent} Recursively back-propagate scores using Eq.~\eqref{eq:backprop}      
    \EndProcedure
    \Procedure{GOBIGraph}{f, G, S, W}
        \State Initialize $i = 0$
        \State \textbf{do}
        \State \hspace{\algorithmicindent} $S \gets S + \gamma \cdot \nabla_S f(G, W, S)$
        \State \hspace{\algorithmicindent} $i \gets i + 1$
        \State \textbf{while} $|\nabla_S f(G, W, S)| > \epsilon$ and $i \leq L$
        \State \textbf{return} $S$
    \EndProcedure
    \Procedure{MCDS}{scheduling interval $I_t$}
        \State $\textsc{MCTS-Test}(root)$
        \State $S_i \gets \textsc{Update}(root)$
        \State Tune $f$ with the new datapoint using~\eqref{eq:loss} as loss \label{line:tune}
        \State \textbf{return} $S_i$
    \EndProcedure
    \end{algorithmic}
\caption{MCDS Scheduler}
\label{alg:testing}
\end{algorithm}
Now that we have trained the deep surrogate model, for an input $[G, W, S]$, we can run the recently developed technique of gradient-optimization using propagation to input (GOBI)~\cite{tuli2021cosco} to update $S$ to maximize the output $f(G, W, S)$ using gradient-ascent rule
\begin{equation}
    S \gets S + \gamma \cdot \nabla_S f(G, W, S).
\end{equation}
    Using advances like momentum, annealing and restarts, we can reach near global optima quickly~\cite{tuli2021cosco}. To differentiate our workflow-graph based surrogate model from the vanilla feed-forward neural network, we call our optimization strategy as \textsc{GOBIGraph} that returns an action in the form of a scheduling decision (see $\textsc{GOBIGraph}$ function in Alg.~\ref{alg:testing}). 
At test time we build the Monte Carlo Tree with slight modifications. In the selection operation, we choose the node using the rule
\begin{align}
\begin{split}
\label{eq:selection_test}
    \text{selected node} &= \argmax_i v_i + \sqrt{\frac{c_1 \times \ln{n}}{n_i}} \\ 
    +\ c_2 \times &\mathbbm{1}(S_i = \textsc{GOBIGraph}(f, G, S, W)).
\end{split}
\end{align}
Here, the third term uses the deep surrogate model to bias our selection choice, like a domain knowledge term commonly used in machine learning~\cite{silver2009monte}. $c_2\in [0, 1]$ is the bias parameter. When performing the simulation operation, we use the value estimate $v = f(G, W, S)$ instead of running expensive co-simulations, reducing the overall scheduling time (line~\ref{line:simulation2} in Alg.~\ref{alg:testing}). \blue{Using the new datapoint in each interval, we also fine-tune the deep surrogate model $f$ in MCDS to adapt in dynamic settings (line~\ref{line:tune} in Alg.~\ref{alg:testing}).}

\section{Performance Evaluation}
\label{sec:perf_eval}
We now describe how we evaluate the MCDS method and compare it against the state-of-the-art baselines: ESVR, DNSGA, IMPSO and Closure as described in Section~\ref{sec:related_work}. We also add a vanilla GOBI method as a reference point for our results to compare against the state-of-the-art deep surrogate model based scheduler.

\subsection{Evaluation Setup}

We use two-test beds to test the scheduling strategies: physical setup and a simulated platform. The former allows us to perform a precise comparison closer to the real-deployments, while the latter lets us perform large-scale experiments. This is a standard testing mechanism in prior work~\cite{tuli2021cosco, basu2019learn}.

\begin{figure}[t]
    \centering \setlength{\belowcaptionskip}{-10pt}
    \includegraphics[width=\linewidth]{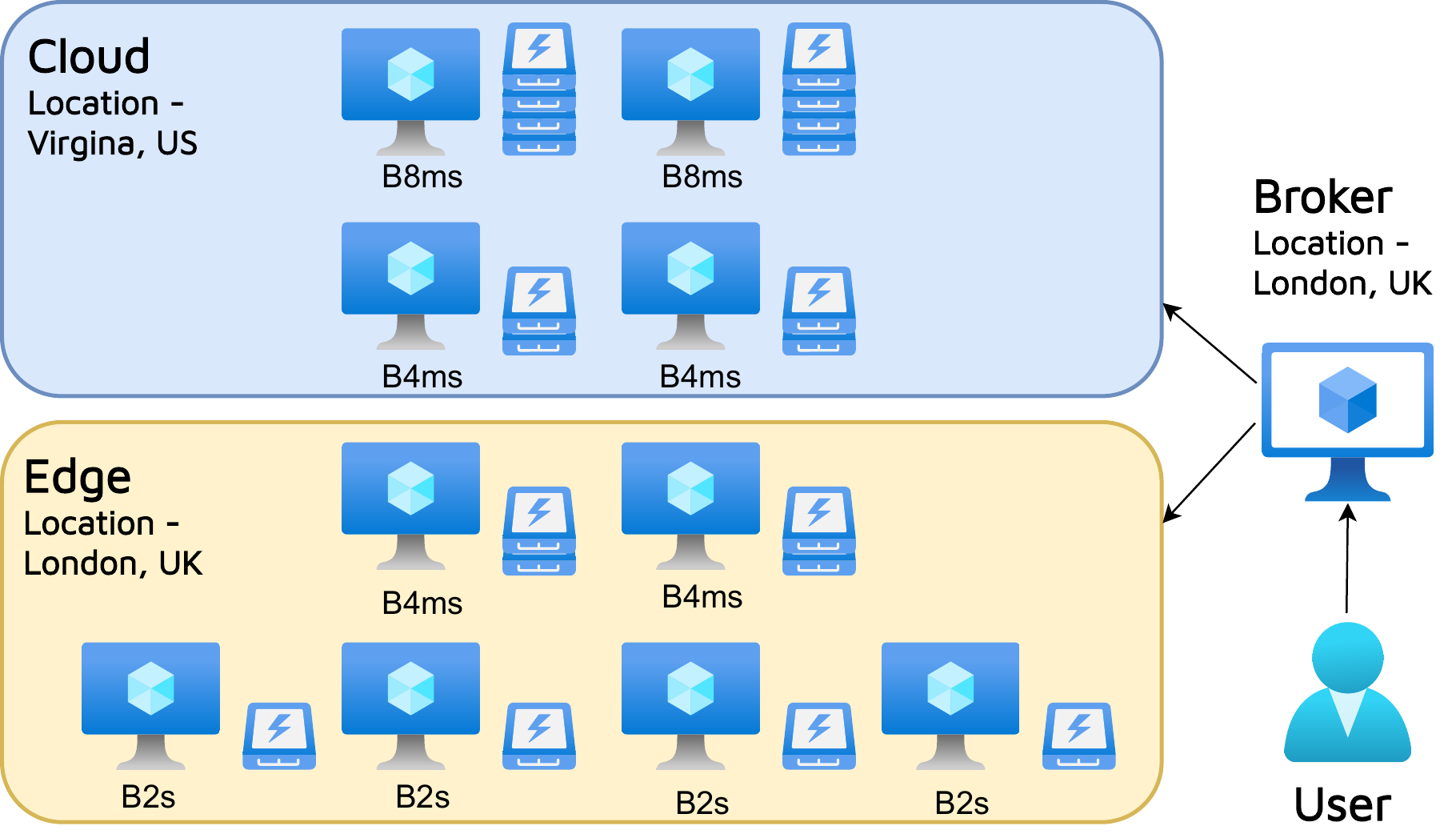}
    \caption{Evaluation Setup.} \vspace{-10pt}
    \label{fig:setup}
\end{figure}

\textbf{Physical Testbed:} We use Microsoft Azure virtual machines (VMs) to create a heterogeneous PDC edge-cloud testbed\footnote{Azure general purpose B-type VMs: \url{https://docs.microsoft.com/en-us/azure/virtual-machines/sizes-b-series-burstable}}. \blue{Specifically, we use the B2s. B4ms and B8ms machine types.} The configuration consists of 10 VMs, 6 light-weight VMs on the edge and 4 powerful ones on a geographically distant cloud location (see Figure~\ref{fig:setup}). All machine \blue{types} use \blue{a different number of} Intel Haswell 2.4 GHz E5-2673 v3 processor cores. To keep storage costs consistent, we keep Azure P15 Managed disk with 125 MB/s disk throughput and 256 GB size\footnote{Azure Managed Disk types \url{https://docs.microsoft.com/en-us/azure/virtual-machines/disks-types\#premium-ssd}}. We use the Microsoft Azure pricing calculator to obtain the cost of execution per hour (in US Dollars)\footnote{Microsoft Azure pricing calculator for South UK \url{https://azure.microsoft.com/en-gb/pricing/calculator/}}. The power consumption models are taken from the SPEC benchmarks repository\footnote{Standard Performance Evaluation Corporation (SPEC) benchmark repository \url{https://www.spec.org/cloud\_iaas2018/results/}}. The Million-Instruction-per-Second (MIPS) of all VMs are obtained using the \texttt{perf-stat}\footnote{\texttt{perf-stat} tool \url{https://man7.org/linux/man-pages/man1/perf-stat.1.html}} tool.

\blue{The Azure B2s machines consist of two cores (4029 MIPS), B4ms have four (8102 MIPS) and B8ms have eight cores (1601 MIPS) each. Moreover, we have 4GB, 16GB and 32GB RAM sizes in B2s, B4ms and B8ms machines, respectively. In our settings, we have considered the response time of edge nodes to be 1 ms and that of cloud nodes to be 10 ms based on the empirical studies using the Ping utility in an existing edge-cloud framework, namely FogBus~\cite{tuli2019fogbus}.} To factor in the possible mobility of the edge nodes, we use the \texttt{NetLimiter} tool to tweak the communication latency with the broker node using the mobility model described in~\cite{tuli2020dynamic}. \blue{Specifically, for the edge nodes in our setup, we use the latency and bandwidth parameters of hosts from the traces generated using the Simulation of Urban Mobility (SUMO) tool~\cite{krajzewicz2012recent} that emulates mobile vehicles in a city like environment. SUMO gives us the parameters like \texttt{ping} time and network bandwidth to simulate in our testbed using \texttt{NetLimiter}.} The edge-cloud broker node was a system situated in {London, UK} and with configuration: Intel i7-10700K CPU, 64GB RAM, and Windows 11 OS.

\textbf{Simulated Testbed:} We consider 50 host machines as a scaled up version of the physical setup. Here, each category has 5 times the instance count, \textit{i.e.}, 30 edge nodes and 20 cloud nodes. This is to test models in a larger-scale edge-cloud environment as considered in prior art~\cite{basu2019learn, ahmed2018docker}. We use a trace driven simulation model that emulates real execution of tasks, but without actual job execution, allowing us to perform experiments on scale.

We run all experiments for 200 scheduling intervals, with each interval being 300 seconds long, giving a total experiment time of 16 hours 40 minutes. We average over five runs and use diverse workload types to ensure statistical significance in our experiments. 

\subsection{Workloads}

For our physical setup, we use \textit{WFCommons} benchmark applications~\cite{coleman2021wfcommons} which are derived from Pegasus workflows~\cite{bharathi2008characterization}. These include the popular non-preemptive scientific applications: BLAST, Cycles and Montage, each having 15 to 100 \blue{tasks sampled uniformly at random}. The BLAST workflow is a high-throughput compute-intensive application for genome based bioinformatics calculations \blue{and has applications in smart health monitoring systems at the edge~\cite{rafique2020complementing}}. The Cycles benchmark performs several agro-ecomonic modelling based calculations \blue{and is popular in edge computing based smart-agriculture~\cite{o2019edge}}. Montage is a benchmark derived from astronomical imaging systems \blue{and similar to other edge computing based space exploration applications~\cite{semenov2021elastic}}. \blue{The Pegasus workflows are popular benchmark applications that are representative of a large number of edge and cloud computing environments. These have been used in several previous works that test workflow scheduling in edge-cloud environments~\cite{lin2019time, shao2019cost, xie2019novel}. Due to these benchmarks being standard in most prior work, we use them in our experiments.}

For our simulator, we use the execution traces of the three applications and the precedence graphs from the benchmark specifications. These traces are generated by running the experiments on the 10 node physical setup. The traces consist of data intervals of 5 minutes, including the CPU, RAM, Disk and Bandwidth utilization of all tasks. We sample monitoring data uniformly from the three applications as done in the \textit{WFCommons's} workload generator.

At the beginning of each scheduling interval we create $Poisson(\lambda)$ workflows with $\lambda = 1.2$ workflows for physical and $\lambda = 5$ workflows for the simulated setup~\cite{tuli2021cosco}. \blue{These values have been taken directly from traces generated from real-world computing applications deployed on Bitbrains datacenters with 10 and 50 servers~\cite{shen2015statisticalBitBrain}. The 1.2 and 5 values have been obtained after normalizing the computational requirements of tasks using the total instructions executed for each task, calculated as the product of average MIPS and execution time. This is done to ensure that the experimental testbed is under a similar load as typical real-world physical deployment.}

\begin{figure*}
    \centering 
    \subfigure[Average Energy Consumption]{
    \includegraphics[width=.235\textwidth]{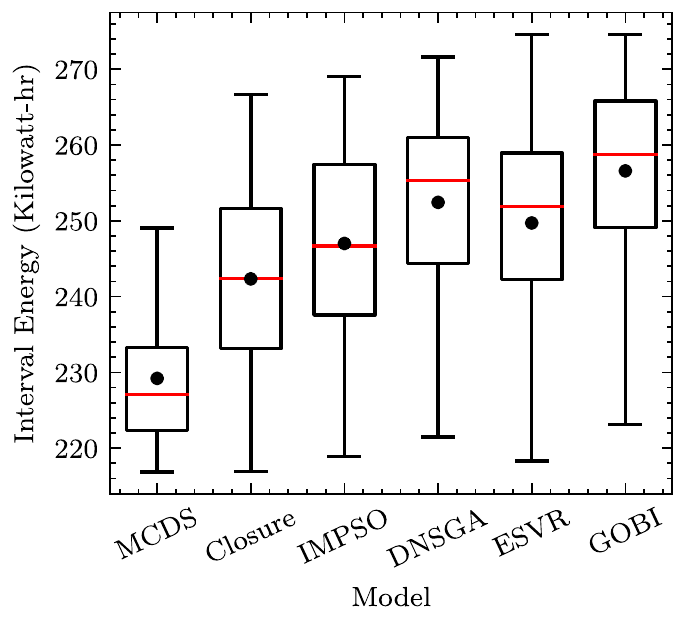}
    \label{fig:f_energy}
    }
    \subfigure[Average Execution Time]{
    \includegraphics[width=.235\textwidth]{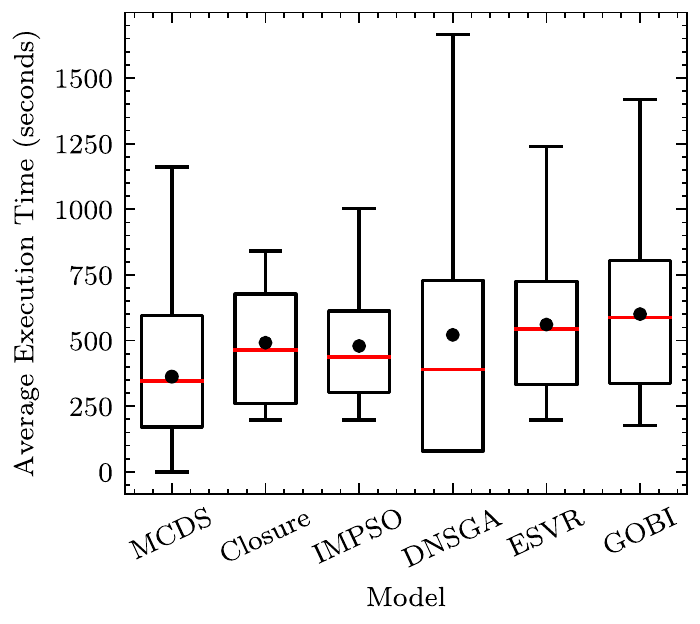}
    \label{fig:f_exec_time}
    }
    \subfigure[Average Wait Time]{
    \includegraphics[width=.235\textwidth]{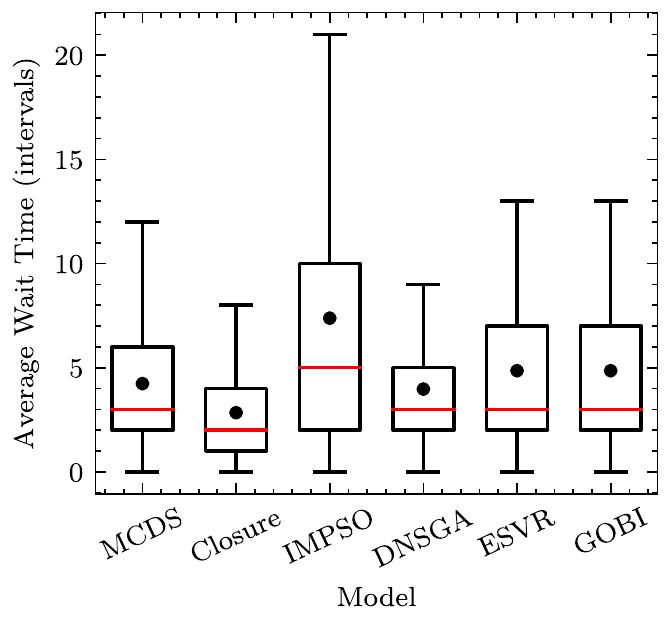}
    \label{fig:f_wait_time}
    }
    \subfigure[Scheduling Time]{
    \includegraphics[width=.235\textwidth]{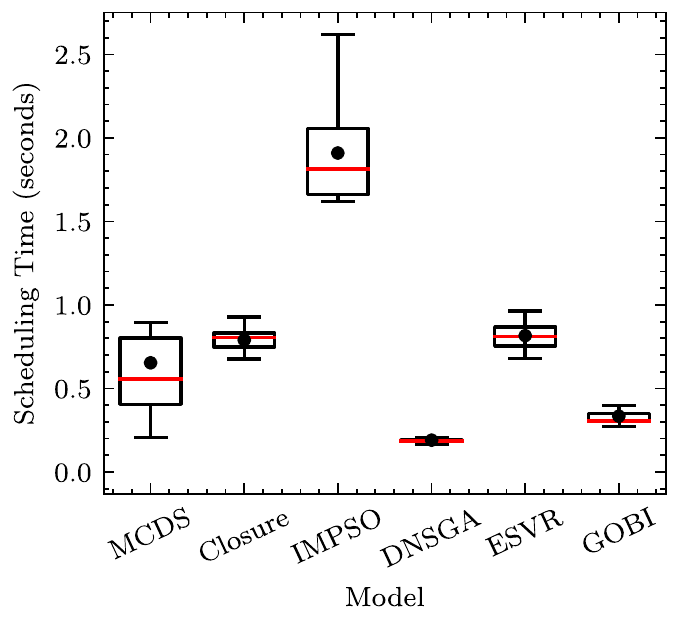}
    \label{fig:f_sched_time}
    }\\
    \subfigure[Average Response Time]{
    \includegraphics[width=.235\textwidth]{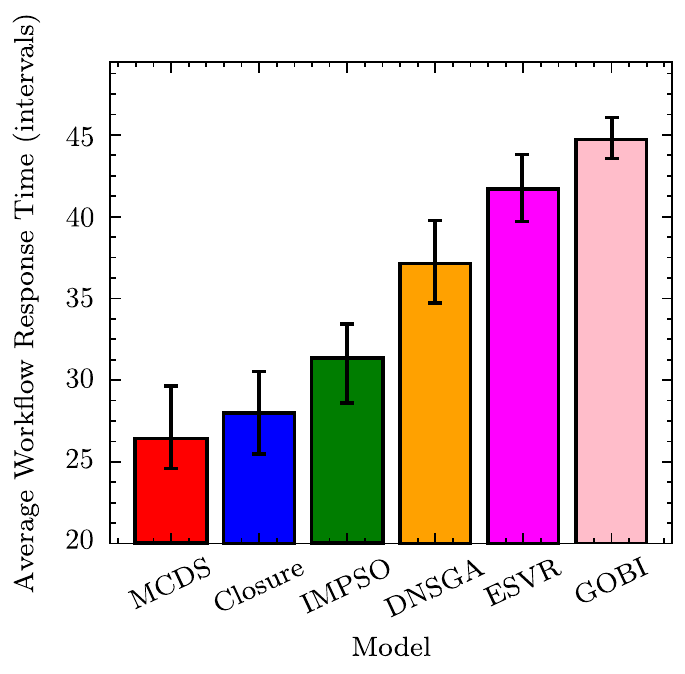}
    \label{fig:f_response}
    }
    \subfigure[Average Response Time (per application)]{
    \includegraphics[width=.235\textwidth]{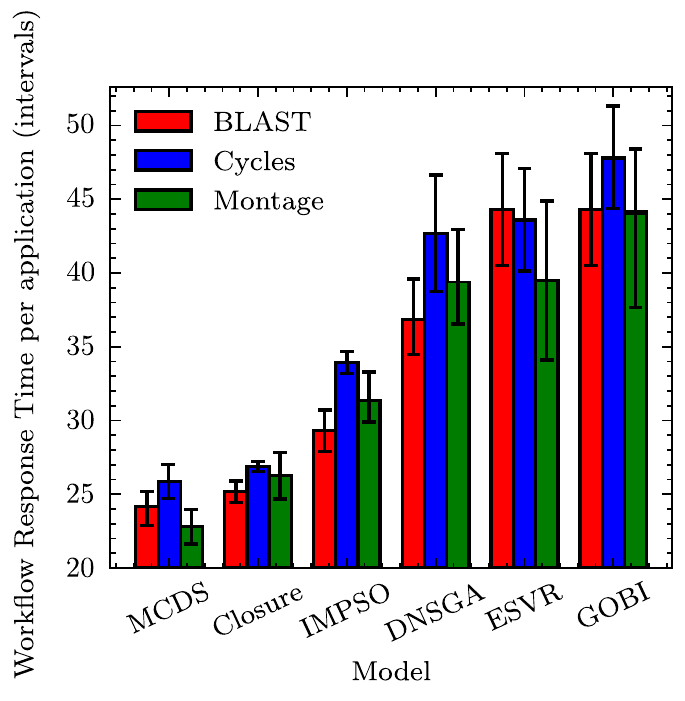}
    \label{fig:f_response_pa}
    }
    \subfigure[SLA Violations]{
    \includegraphics[width=.235\textwidth]{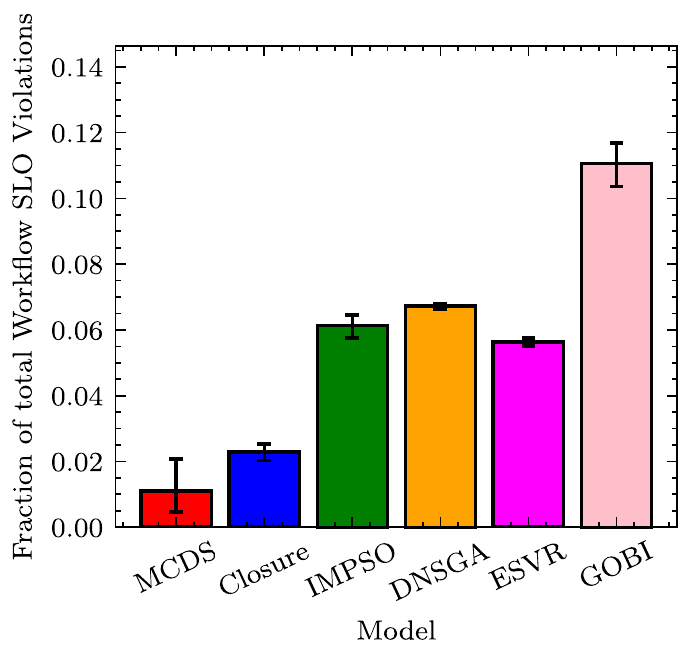}
    \label{fig:f_sla}
    }
    \subfigure[SLA Violations (per application)]{
    \includegraphics[width=.235\textwidth]{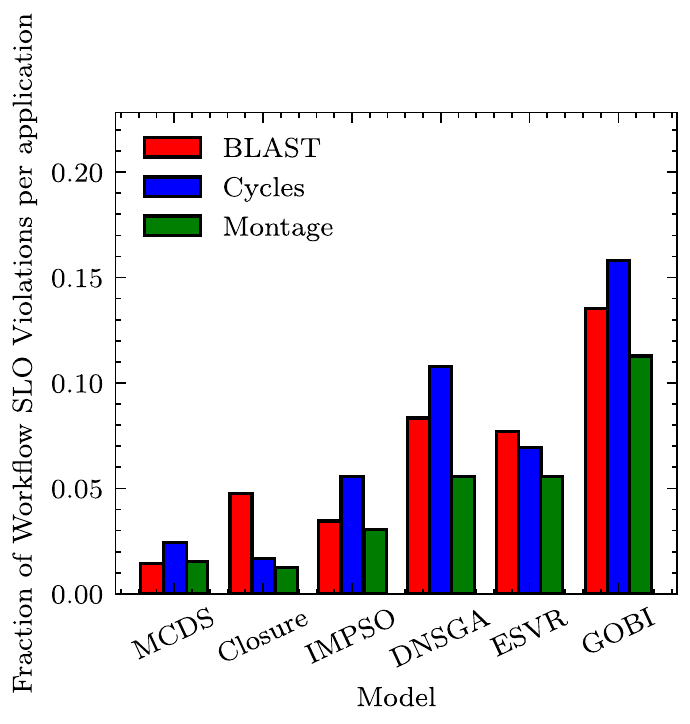}
    \label{fig:f_sla_pa}
    }\\
    \subfigure[Fairness]{
    \includegraphics[width=.235\textwidth]{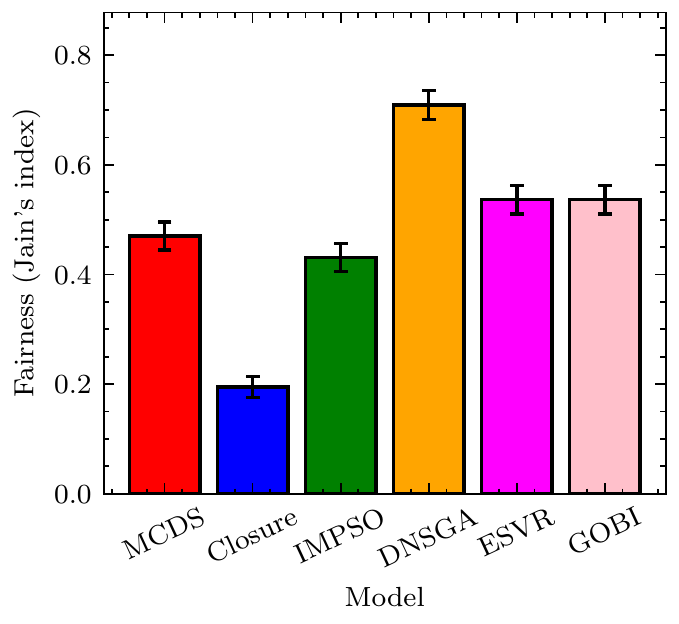}
    \label{fig:f_fairness}
    }
    \subfigure[Cost per Workflow]{
    \includegraphics[width=.235\textwidth]{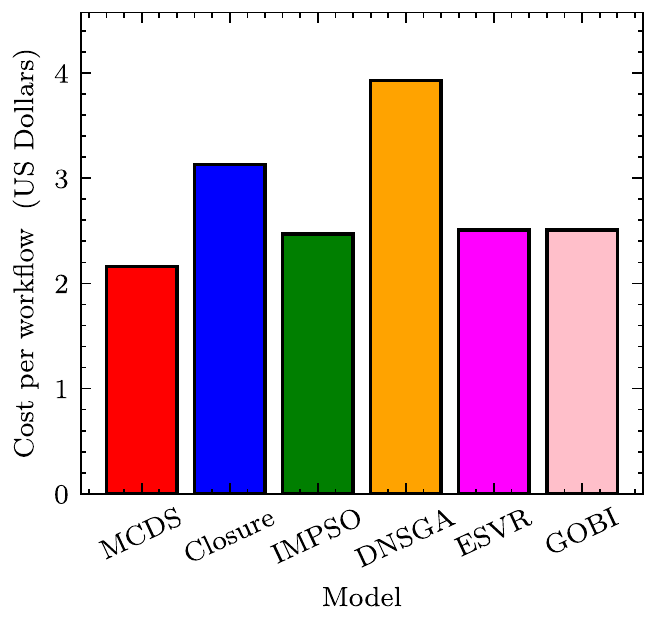}
    \label{fig:f_cost}
    }
    \subfigure[Migration Time with intervals]{
    \includegraphics[width=.235\textwidth]{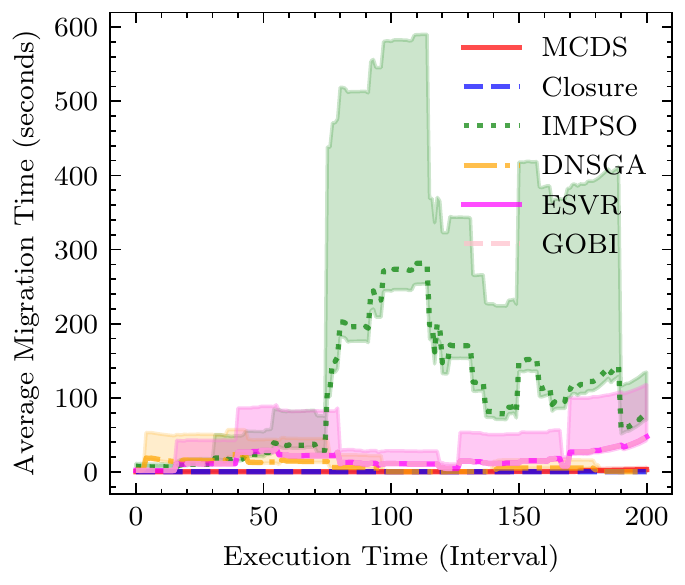}
    \label{fig:f_migration_time}
    }
    \subfigure[Number of Task Migrations with intervals]{
    \includegraphics[width=.235\textwidth]{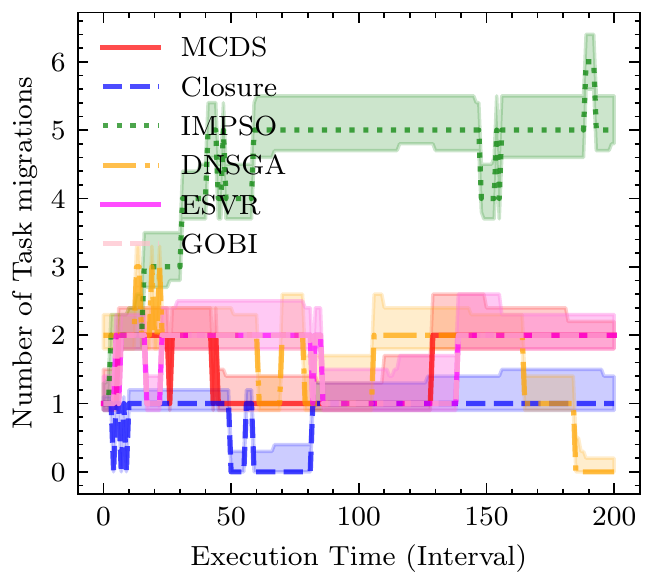}
    \label{fig:f_num_migrations}
    }
    \caption{Comparison of MCDS against baselines on physical setup with 10 hosts. Box plots: dot represents mean and solid red line represents median. Error bars in the bar plots correspond to $90\%$ confidence-interval. Shaded regions in the line plots show the range of the values over five runs.} \vspace{-10pt}
    \label{fig:framework_results}
\end{figure*}

\begin{figure*}
    \centering \setlength{\belowcaptionskip}{-12pt}
    \subfigure[Average Energy Consumption]{
    \includegraphics[width=.235\textwidth]{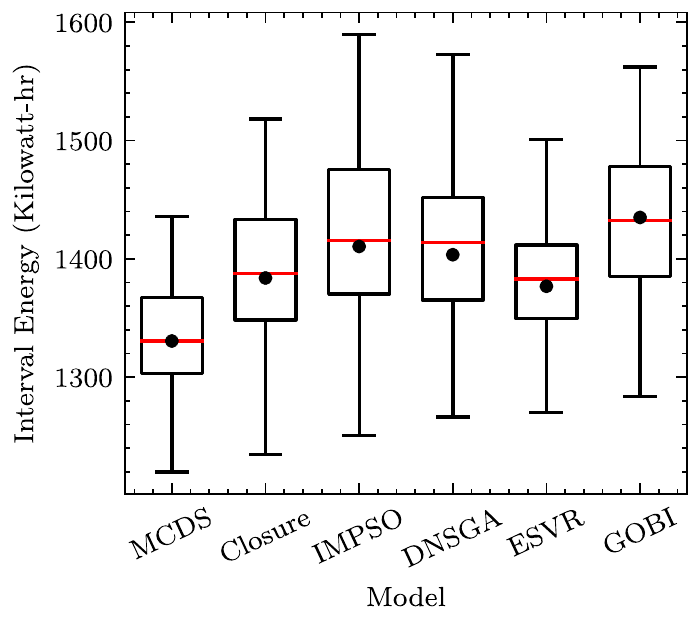}
    \label{fig:s_energy}
    }
    \subfigure[Average Execution Time]{
    \includegraphics[width=.235\textwidth]{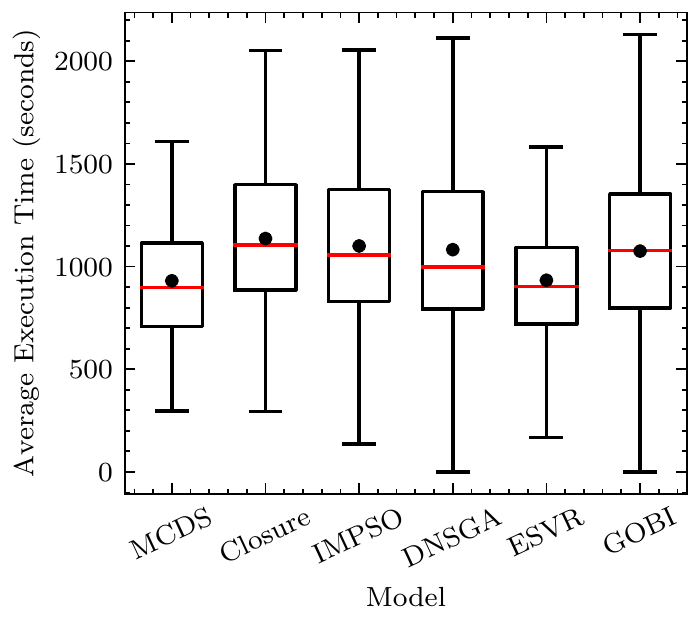}
    \label{fig:s_exec_time}
    }
    \subfigure[Average Wait Time]{
    \includegraphics[width=.235\textwidth]{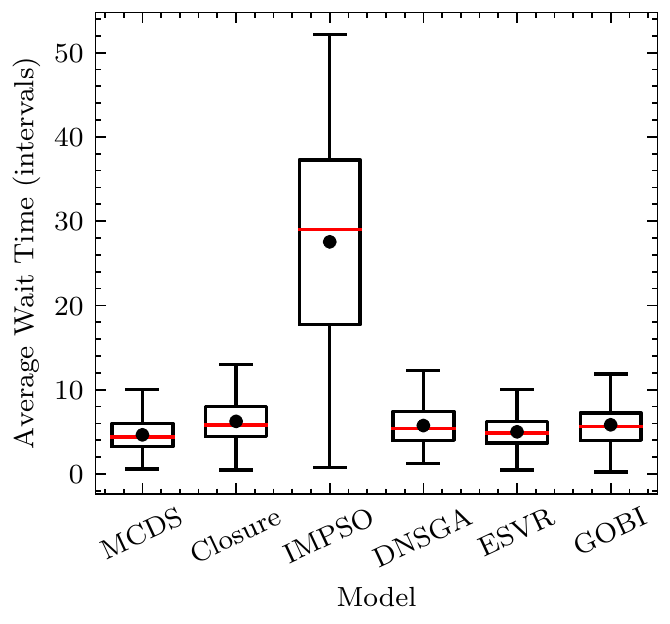}
    \label{fig:s_wait_time}
    }
    \subfigure[Scheduling Time]{
    \includegraphics[width=.235\textwidth]{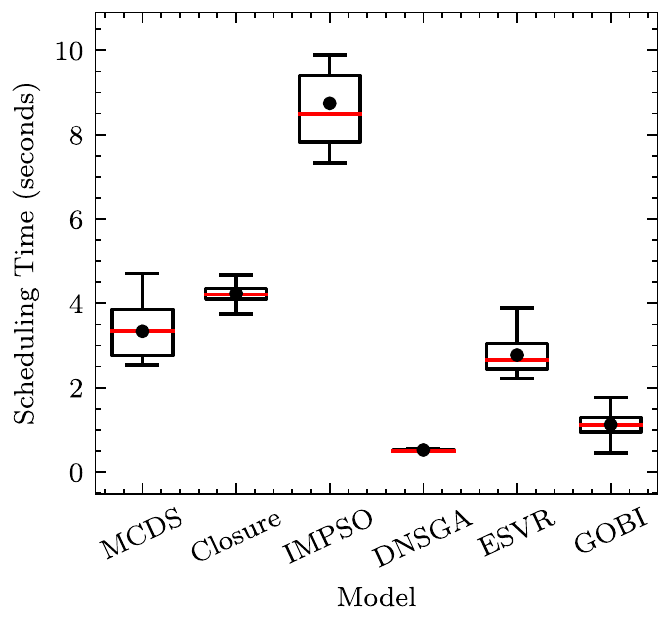}
    \label{fig:s_sched_time}
    }\\
    \subfigure[Average Response Time]{
    \includegraphics[width=.235\textwidth]{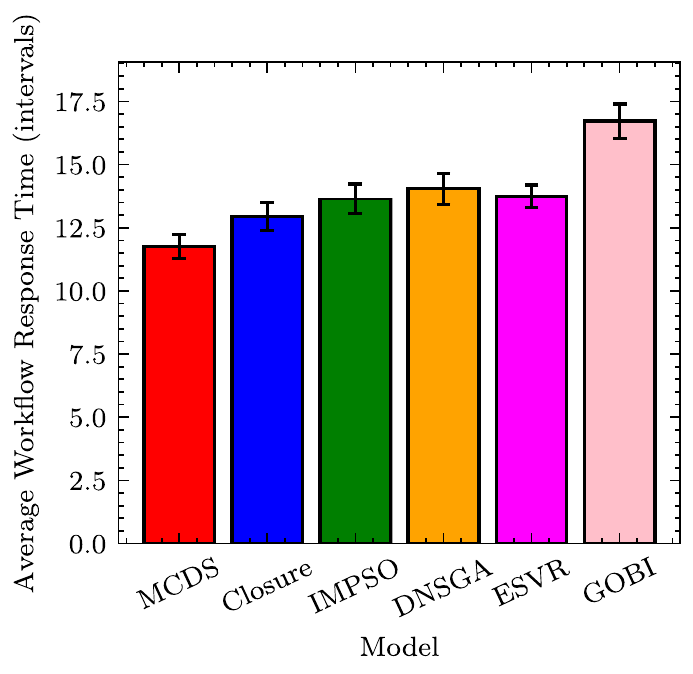}
    \label{fig:s_response}
    }
    \subfigure[Average Response Time (per application)]{
    \includegraphics[width=.235\textwidth]{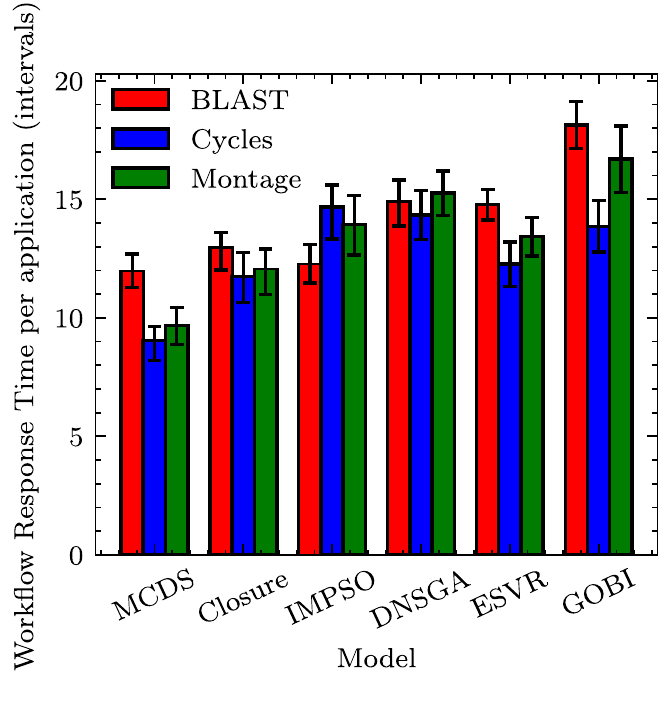}
    \label{fig:s_response_pa}
    }
    \subfigure[SLA Violations]{
    \includegraphics[width=.235\textwidth]{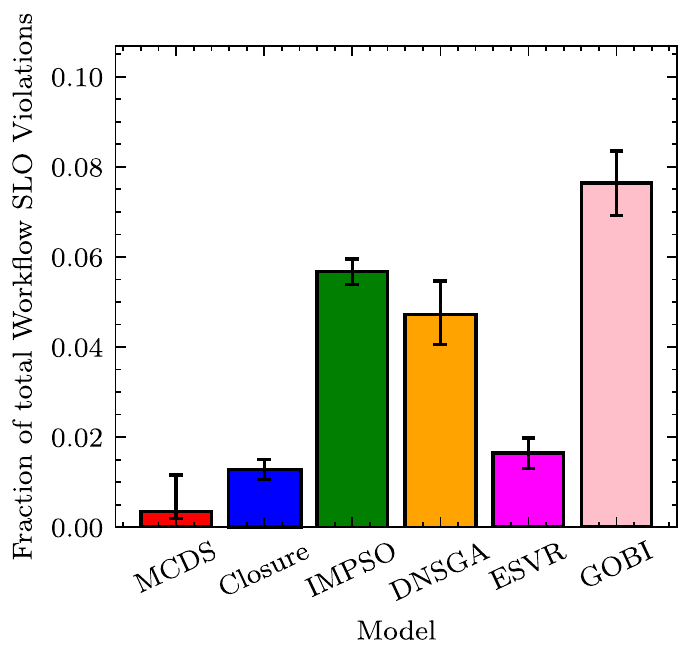}
    \label{fig:s_sla}
    }
    \subfigure[SLA Violations (per application)]{
    \includegraphics[width=.235\textwidth]{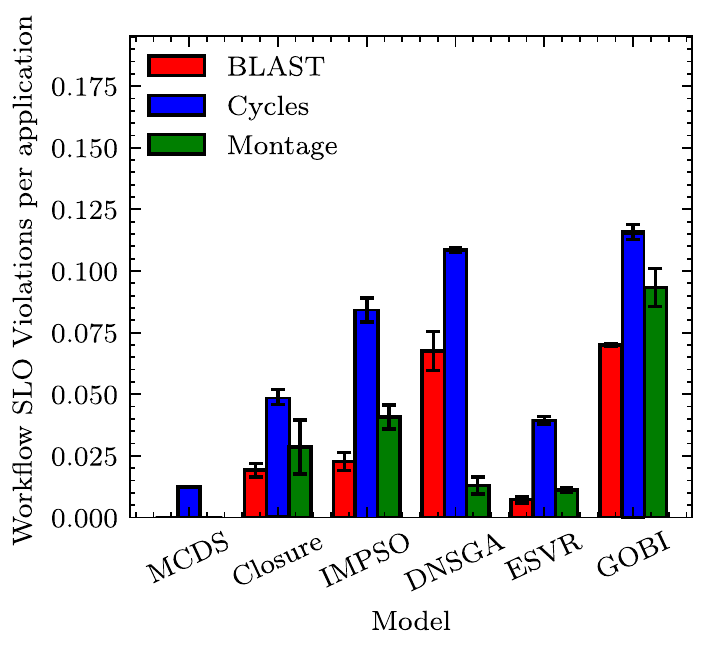}
    \label{fig:s_sla_pa}
    }\\
    \subfigure[Fairness]{
    \includegraphics[width=.235\textwidth]{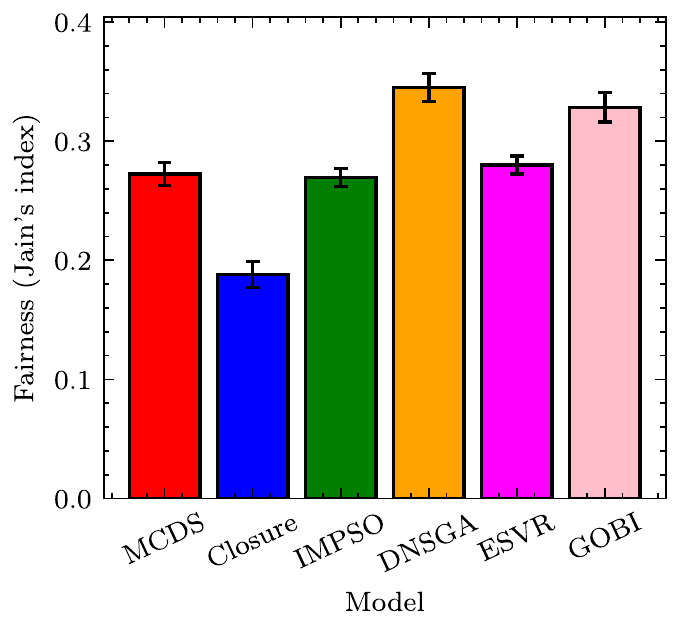}
    \label{fig:s_fairness}
    }
    \subfigure[Cost per Workflow]{
    \includegraphics[width=.235\textwidth]{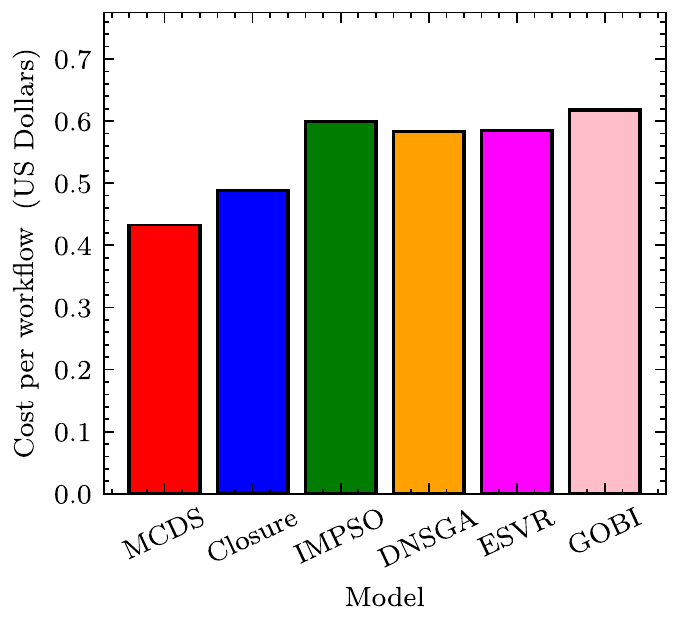}
    \label{fig:s_cost}
    }
    \subfigure[Migration Time with intervals]{
    \includegraphics[width=.235\textwidth]{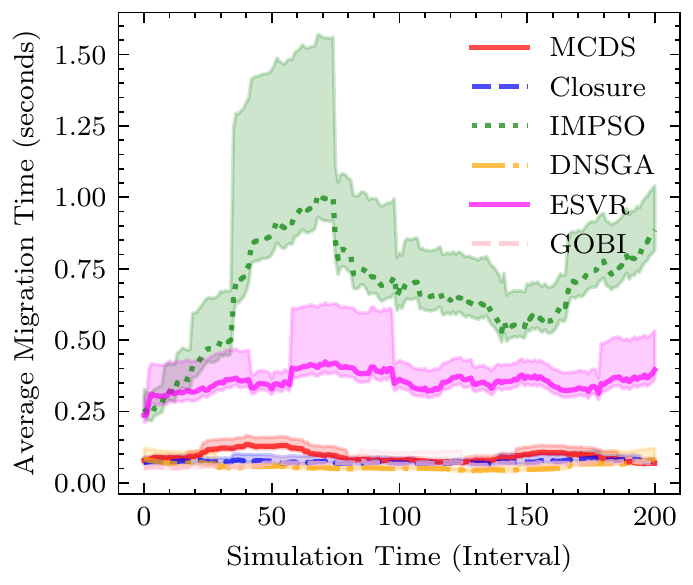}
    \label{fig:s_migration_time}
    }
    \subfigure[Number of Task Migrations with intervals]{
    \includegraphics[width=.235\textwidth]{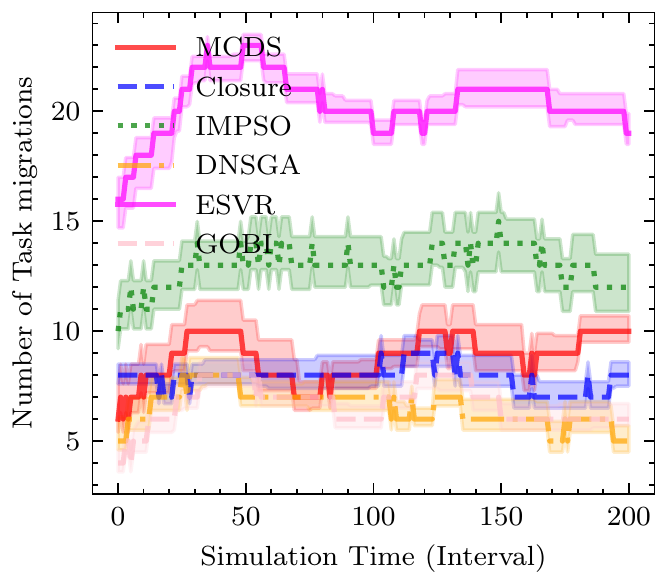}
    \label{fig:s_num_migrations}
    }
    \caption{Comparison of MCDS against baselines on simulated setup with 50 hosts.}
    \label{fig:simulator_results}
\end{figure*}

\subsection{Implementation Details}

We perform all our experiments on an extended version of the COSCO framework~\cite{tuli2021cosco} to support workflow DAG models. COSCO~\cite{tuli2021cosco} is a framework for the development and deployment of applications on edge-cloud environments with structured communication and platform independent execution of applications. COSCO connects all edge and cloud nodes in a physical setup using HTTP REST APIs. We deploy our workflows in COSCO, with each task mapped to a docker container.

To implement MCDS in the COSCO framework, we extend the \texttt{Framework} class. The function \texttt{getPlacementPossible()} was modified to also check for precedence constraints when allocating tasks. Moreover, we implemented a data transferring pipeline for broadcasting inputs for parallel tasks and forwarding the outputs between two neighboring tasks in a DAG. Finally, the outputs were synchronized and brought to the broker to calculate the inference accuracy and measure the workflow response time. We use the HTTP Notification API to synchronize outputs and execute network splits. Further, the GOBIGraph approach is implemented using the Autograd package in the PyTorch module~\cite{paszke2017automatic}.

\blue{To run an experiment, the broker runs the COSCO framework and communicates with each host via HTTP REST APIs~\cite{yates2015ensembl}. The REST server of the \texttt{Framework} class sends operation codes including \texttt{GetHostStats, GetContainerStats, Checkpoint, Migrate} and \texttt{Restore} to each host to instruct a specific activity. At every worker host, a \texttt{DockerClient} service runs with a \texttt{Flask} HTTP web-service~\cite{grinberg2018flask}. Flask is a web application REST client that allows interfacing between edge-cloud hosts and the broker.  To obtain the resource utilization metrics of the host machines at the start of each scheduling interval, we use the \texttt{psutil} utility for memory and disk characteristics, \texttt{perf stat} for CPU utilization and \texttt{ioping} utility for memory and network bandwidth utilization characteristics. For the running tasks, we use the \texttt{docker inspect} command with the identifier of the corresponding container instance as an argument. All utilization metrics are instantly saved to an InfluxDB database instance running at the broker node. This additional step of collecting utilization metrics leads to an overhead due to the HTTP REST request and adding a new entry in the database. However, this occurs in a non-blocking fashion, \textit{i.e.}, it does not interrupt the execution of active containers in worker hosts.}

\subsection{Model and Experimental Assumptions}
To optimize the QoS parameters, we consider a standard objective function that focuses on energy consumption and response time as is done in prior work~\cite{tuli2021cosco}. For interval $I_t$,
\begin{equation}
    \label{eq:objective_function}
    \mathcal{O}_t = 1 - \alpha \cdot AEC_t - (1-\alpha) \cdot ART_t,
\end{equation}
where $AEC_t$ and $ART_t$ are the normalized average energy consumption of the environment in $I_t$ and average response time of workflows leaving at the end of $I_t$. Here, $\alpha$ is a hyper-parameter that can be set by users as per the application requirements. \blue{We choose response time as a metric to optimize the service quality by reducing the latency of the executed workflow applications and subsequently the SLA violation rates~\cite{gill2019transformative}. Energy is another important metric, both for edge and cloud nodes. At the edge, devices are typically mobile, relying on battery operated hardware. For long-lasting services and reducing the average down-times of edge nodes, supreme energy efficiency is crucial. For cloud service providers, higher energy consumption leads to increased operation expenses, both for running the cloud servers as well as the cooling equipment~\cite{tuli2021hunter}. The $\alpha$ parameter is chosen the user as per the application demands.}

For our experiments, we keep $\alpha = 0.5$ in~\eqref{eq:objective_function} \blue{as it is a characteristic representative of typical fog environments~\cite{tuli2021cosco, tuli2020dynamic}.} We also set $\phi = 10$ (number of actions in simulations for model training) and $\psi = 50$ (number of MCTS iterations). We also use $c_1 = 0.1, c_2 = 0.5$. All hyper-parameter values are obtained using grid-search.  We get similar trends for other values. We present a sensitivity analysis on $\psi$ later in Section~\ref{sec:sens}. 
To train our deep surrogate model, we use we the AdamW optimizer~\cite{adamw} with learning rate as $0.0001$.

\subsection{Evaluation Metrics}
\label{sec:eval_metrics}

To compare the QoS, we consider metrics like energy consumption and response time. We also compare the SLA violation rates. The SLA of a workflow is violated if its response time is greater than the deadline. We consider the relative definition of SLA (as in~\cite{tuli2021cosco}) where the deadline is the 98$^{th}$ percentile response time for the same application (BLAST/Cycles/Montage) on the state-of-the-art baseline \textit{Closure}. \blue{To ensure sufficient sample size, we run the \textit{Closure} baseline for 1000 scheduling intervals and evaluate the 98$^{th}$ percentile response times for each application.} We also consider scheduling fairness, for which we use Jain's fairness index over the IPS of the running tasks in the system. We also compare the average migration time, which is the average time for task (container) migrations.
We also compare the scheduling time, execution cost and average waiting time (number scheduling intervals spent in the wait queue) of all executed workflows.

\subsection{Results}

We show the results of our experiments on the physical and simulated testbeds in Figures~\ref{fig:framework_results} and~\ref{fig:simulator_results} respectively. Figures~\ref{fig:f_energy} and~\ref{fig:s_energy} show that the MCDS model has the lowest energy consumption compared to the baseline methods. MCDS scheduler gives an average interval energy consumption of 228.21 KW-hr for 10 hosts and 1326.98 KW-hr for 50 hosts. As the number of hosts increases so does the power consumption of the setup. However, MCDS has up to 6.13\% lower energy consumption than the second lowest value of the Closure scheduler. \blue{The major reason for this is the low response time of each application, which leads to lower number of active tasks in the system at any point giving low CPU utilization and consequently low energy consumption.} Figures~\ref{fig:f_exec_time} and~\ref{fig:s_exec_time} show the execution times of the workflows, \textit{i.e.}, the time these workflows spend in computation. All schedulers have similar execution times for both setups. Figures~\ref{fig:f_wait_time} and~\ref{fig:s_wait_time} show the waiting times of workflows. Here, IMPSO has the highest waiting times as using this scheduler leads to the maximum number of resource contentions in the system. The scheduling times of all schedulers are less than 10 seconds (3.33\% of the scheduling interval duration of 5 minutes), with DNSGA and GOBI having the lowest scheduling times (Figures~\ref{fig:f_sched_time} and~\ref{fig:s_sched_time}). MCDS has 13.19\% lower scheduling time than the best performing baseline, \textit{i.e.}, Closure. \blue{A higher scheduling time compared to the meta-heuristic approach DNSGA is due to the multiple Monte-Carlo iterations. However, the model is faster than other approaches like Closure and IMPSO, showing that MCDS is scalable to larger testbeds. More details on scalability with workload size in Section~\ref{sec:scale}.}

Figures~\ref{fig:f_response} and~\ref{fig:f_response} show the average response times for both setups. Among the baselines, Closure has the lowest average response time of 27.21 intervals in the physical setup and 12.67 intervals in the simulated case (lower response time in the simulated setup due to the larger parallelization capacity). MCDS has a response time up to 4.56\% lower than Closure. \blue{This is due to the long-term optimization in MCDS, unlike other methods. Myopic optimization in prior work typically prevents them from considering future system states that could cause an increase in response time due to resource contention or migrations. This allows MCDS to be parsimonious in terms of number of preemptive migrations (Figs.~\ref{fig:f_num_migrations} and~\ref{fig:s_num_migrations}). Comparing the number of migrations across baselines, we see that IMPSO and ESVR aggressively migrate tasks, whereas the other methods try to minimize migration overheads to optimize QoS. } Low response times of the MCDS scheduler also results in low SLA violation rates (here the service level objectives are defined as deadlines as discussed in Section~\ref{sec:eval_metrics}) as seen in Figures~\ref{fig:f_sla} and~\ref{fig:s_sla}. MCDS has an SLA violation percentage of 0.25\%-1.12\%, lower than the best values among the baselines, which are 1.12\%-2.04\% for the Closure scheduler. IMSO has a high SLA violation rates due to the high waiting and migration times as seen in Figures~\ref{fig:f_migration_time} and~\ref{fig:s_migration_time}. \blue{This is due to the aggressive migration in IMPSO as seen in Figures~\ref{fig:f_num_migrations} and~\ref{fig:s_num_migrations}.} ESVR has relatively low SLA violation rates for its response times due to the deadline-first prioritization in the scheduling decisions. Figures~\ref{fig:framework_results} and~\ref{fig:simulator_results} also show the response times and SLA violation rates for each application type (BLAST, Cycles and Montage). \blue{The SLA violation rates for each application is shown in Figures~\ref{fig:f_sla_pa} and~\ref{fig:s_sla_pa}. The deviations from the $2\%$ mark in these figures is due to the stochasticity in the workloads, motivating us to fine-tune the model at each interval (line~\ref{line:tune} in Alg.~\ref{alg:testing}).}

Comparing the fairness indices of the schedulers (Figures~\ref{fig:f_fairness} and~\ref{fig:s_fairness}), we see that the DNSGA has the maximum fairness index of 0.34-0.72. MCDS has a fairness index of 0.28-0.52. \blue{This is due to the long-term optimization in MCDS, preventing resource contention and allocation of resources to tasks corresponding to their requirements.} As MCDS is able to complete the maximum number of workloads in the duration of 200 intervals, the amortized cost per workflow of MCDS is lowest among all models (0.43 and 2.21 USD/workflow). MCDS gives 16.12\%-30.71\% lower cost than the baselines with ten hosts (Figure~\ref{fig:f_cost}). Similar trends are seen in the simulated setup with 50 hosts (7.62\%-70.14\%).

\subsection{Ablation Analysis}

To study the relative importance of each component of the model, we exclude every major one and observe how it affects the performance of the scheduler. An overview of this ablation analysis is given in Table~\ref{tab:ablation}. First, we consider the MCDS scheduler without the Monte-Carlo tree search, \textit{i.e.}, only using the pre-trained deep surrogate model to get the GOBIGraph output (\textit{w/o MCTS} model). Second, we consider a model without the deep surrogate model, \textit{i.e.}, we only use MCTS simulations with a random scheduler to get the QoS estimates and generate the scheduling decision at each interval (\textit{w/o DSM} model). The other two models we consider are modifications in the selection formula. In the \textit{w/o Exploration} model we ignore the UCB exploration metric. In the \textit{w/o Domain Kn.} we ignore the GOBIGraph based indicator term in~\eqref{eq:selection_test}. We reach the following findings:

\begin{figure*}[!t]
    \centering \setlength{\belowcaptionskip}{-12pt}
    \includegraphics[width=.7\textwidth]{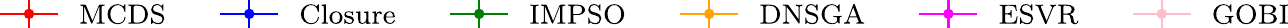}\\
    \subfigure[Energy Consumption]{
    \includegraphics[width=.235\textwidth]{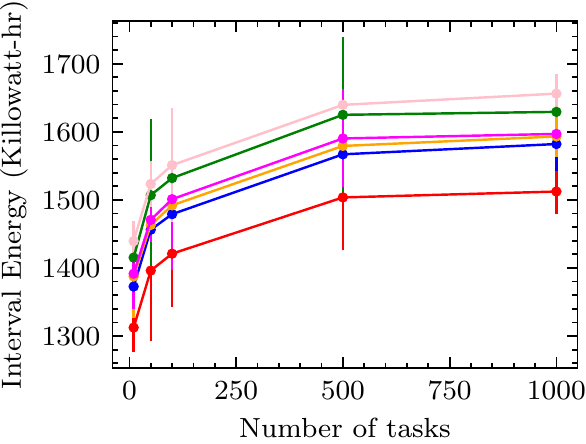}
    \label{fig:scal_energy}
    }
    \subfigure[Response Time]{
    \includegraphics[width=.235\textwidth]{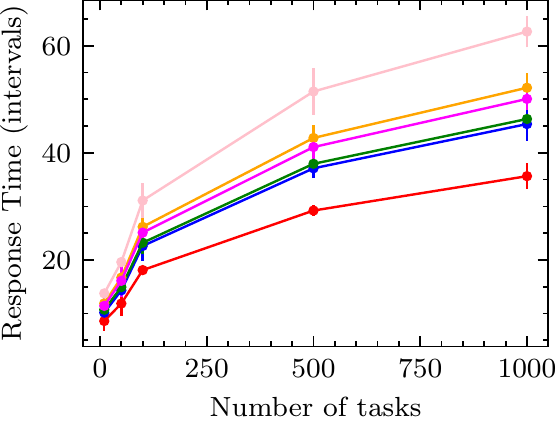}
    \label{fig:scal_response}
    }
    \subfigure[SLA Violations]{
    \includegraphics[width=.235\textwidth]{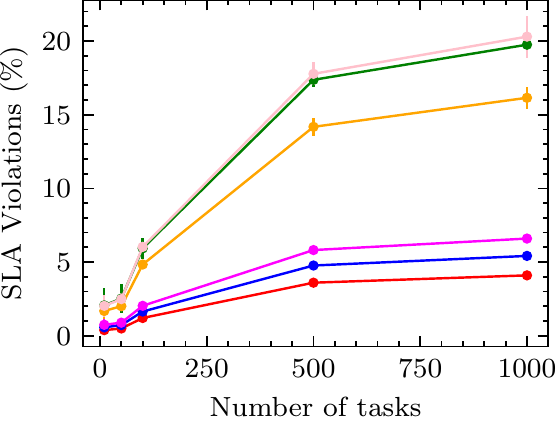}
    \label{fig:scal_sla}
    }
    \subfigure[Scheduling Time]{
    \includegraphics[width=.235\textwidth]{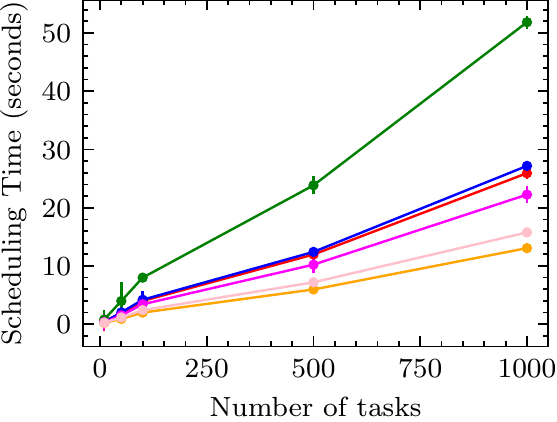}
    \label{fig:scal_sched}
    }
    \caption{\blue{Scalabilty Analysis for MCDS and baseline methods with the number of tasks in each workflow as 10, 50, 100, 500 and 1000.}}
    \label{fig:scal}
\end{figure*}

\begin{figure*}[!t]
    \centering \setlength{\belowcaptionskip}{-12pt}
    \includegraphics[width=.8\textwidth]{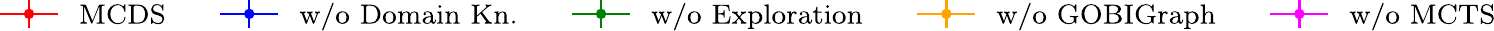}\\
    \subfigure[Energy Consumption]{
    \includegraphics[width=.235\textwidth]{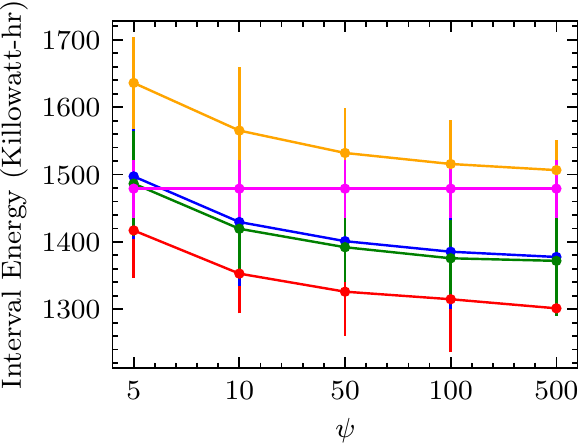}
    \label{fig:sens_energy}
    }
    \subfigure[Response Time]{
    \includegraphics[width=.235\textwidth]{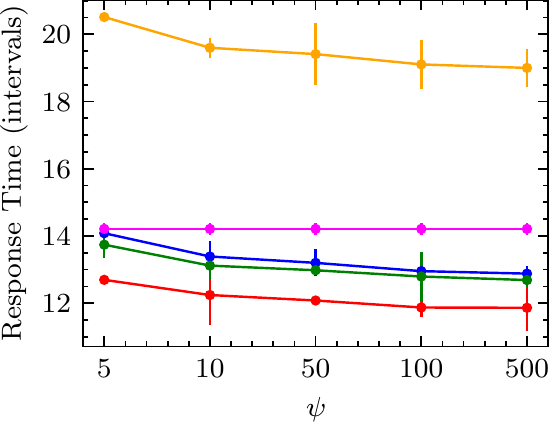}
    \label{fig:sens_response}
    }
    \subfigure[SLA Violations]{
    \includegraphics[width=.235\textwidth]{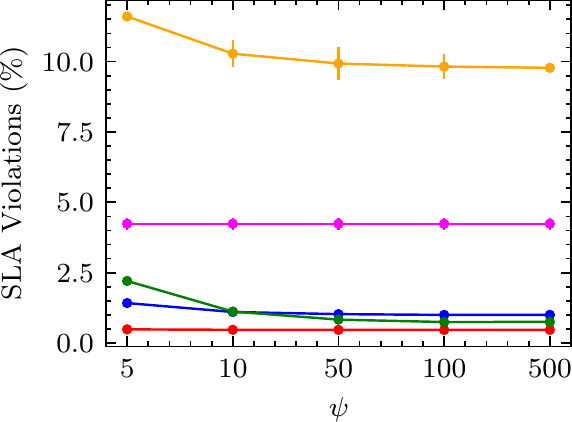}
    \label{fig:sens_sla}
    }
    \subfigure[Scheduling Time]{
    \includegraphics[width=.235\textwidth]{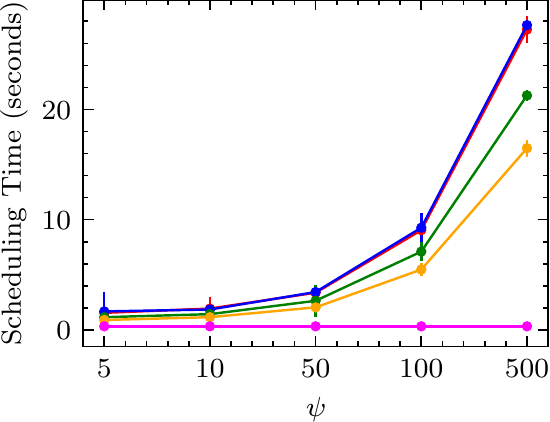}
    \label{fig:sens_sched}
    }
    \caption{Sensitivity Analysis for ablated models with number of Monte Carlo iterations (the $\psi$ parameter in our formulation).}
    \label{fig:sens}
\end{figure*}

\begin{table}[t]
    \centering 
    \caption{Ablation Analysis. Units: Energy - \textit{KW-Hr}, Response Time - scheduling intervals, SLA Violations - percentage, Scheduling Time (cumulative) - seconds.}
    \resizebox{\linewidth}{!}{
    \begin{tabular}{@{}lcccccc@{}}
    \toprule 
    Model & Energy & Resp. Time & SLA Violation & Sched. Time\tabularnewline
     \midrule
    MCDS & \textbf{1326} & \textbf{12.08} & \textbf{0.468} & 3.3627\tabularnewline
    \midrule 
    w/o MCTS & 1479 & 14.21 & 4.283 & \textbf{0.3123}\tabularnewline
    w/o DSM & 1401 & 13.20 & 3.027 & 7.4231\tabularnewline
    w/o Exploration & 1392 & 12.98 & 0.832 & 2.6283\tabularnewline
    w/o Domain Kn. & 1532 & 19.41 & 0.930 & 2.0318\tabularnewline
    \bottomrule 
    \end{tabular}}
    \label{tab:ablation}
\end{table}

\begin{itemize}[leftmargin=*]
    \item Without the MCTS search scheme, the model performs a myopic optimization and suffers the maximum increase in the  SLA violation rate. This demonstrates the need for a long-term objective estimation and bootstrapping is crucial to reach optimal QoS.
    \item Without the deep-surrogate model, the scheduling time significantly increases. This is due to the large number of co-simulation steps being performed in the MCTS iterations. It also affects the QoS values due to the inaccurate estimates obtained from the simulations based on the random scheduler. 
    \item Without the exploration term, the model is unable to find possibly better scheduling decisions. Also, the domain knowledge term, \textit{i.e.}, the GOBIGraph based indicator term in ~\eqref{eq:selection_test} helps the model to use the deep surrogate model for exploration. Both these terms help improve the scheduling decisions and reach better QoS values. 
\end{itemize}
We observe that Monte-Carlo learning and deep surrogate model have the maximum impact on the performance. UCB exploration and domain knowledge are additional features that help improve QoS.

\subsection{Scalability Analysis}
\label{sec:scale}

\blue{The \textit{WfGen} utility within \textit{WFCommons} benchmark suite allows us to change the number of tasks in each workflow DAG.\footnote{\blue{WfGen - WfCommons \url{https://wfcommons.org/generator}.}} In our main experiments, we sample the number of jobs uniformly at random between 15 and 100. We now use this feature to test the scalability of the MCDS model and compare it against the baselines on the simulated setup with 50 hosts. Figure~\ref{fig:scal} shows the energy consumption, response time, SLA violations and scheduling time when we vary the number of tasks in each workflow from 10 to 1000.  Comparing across models, the DNSGA, ESVR and IMPSO do not scale well in terms of SLA violation rates as the number of tasks grows. Only models that use deep surrogate networks, \textit{i.e.}, MCDS and GOBI, and the attack-defense based scalable baseline Closure are able to scale with increasing task count. Similarly, other parameters, \textit{i.e.}, energy consumption, response time and scheduling time increase with the number of tasks. In terms of energy consumption, response time, SLA violation rate and scheduling time, MCDS not only has better metric scores, but also does not face from excessive performance drop in cases where workflows have a large number of tasks, thanks to its gradient graph convolution based adaptive normalization. The deep surrogate model in MCDS is able to handle large-scale workflows also due to the \textit{Bahdanau} attention that brings down the dimensional size of the input size windows. }

\subsection{Sensitivity Analysis}
\label{sec:sens}

We check the sensitivity of the performance parameters to the $\psi$ parameter (number of Monte-Carlo iterations), as shown in Figure~\ref{fig:sens}. The figure shows the energy consumption, response time, SLA violations and scheduling time when $\psi$ varies from 5 to 500. \blue{As the number of iterations increases, the model can cover larger state-action space and get a much better look-ahead estimate.} Thus, the energy consumption, response time and SLA violations decrease as $\psi$ increases (except the \textit{w/o MCTS} model which is unaffected as MCTS is not performed here). Although, after $\psi = 50$ the improvement in the QoS performance is minimal. 
However, increasing the simulation steps has an adverse impact on the scheduling time of the model. We choose $\psi = 50$ in our experiments to balance the trade-off between QoS and scheduling overhead.

\section{Conclusions and Future Work}
\label{sec:conclusions}

We have presented MCDS, a novel scheduler for efficiently scheduling demanding workflow applications on mobile edge-cloud computing environments. Unlike prior work that performs myopic optimization of QoS using reinforcement learning, evolutionary methods or heuristics, MCDS trains a long-term QoS estimation surrogate model. To do this, MCDS uses a modified version of the Monte-Carlo-Tree-Search that uses UCB exploration and gradient-optimization based biased exploitation. This allows MCDS to manage workflows to maintain high inference objective scores on an average and reduce the scheduling time as it does not need to perform several multi-step simulations at test time. All these contributions allow MCDS to \blue{be scalable and} outperform the state-of-the-art models in terms of energy consumption, response time, SLA violation rate and cost by up to 6.13\%, 4.56\%, 45.09\% and 30.71\% respectively in a heterogeneous mobile edge-cloud environment with real-world workloads. 

In some application scenarios, tasks in the same workflow DAG use the same input files for processing. Thus, allocating such tasks on the same host could help reduce data transfer overheads. In the future, we wish to extend the proposed model also to consider data transfer overheads when scheduling workflows. \blue{We also propose to tune the MCDS approach for more recent IoT-based workloads that are based on augmented reality, video processing and Big Data analytics applications~\cite{yi2015survey}.}

\section*{Software Availability}
\footnotesize{The code is available at \blue{\url{https://github.com/imperial-qore/COSCO/tree/workflow}}. The training datasets are available at \url{https://doi.org/10.5281/zenodo.5779005}, released under the CC BY 4.0 license. The Docker images used in the experiments are available at \url{https://hub.docker.com/u/shreshthtuli}. }

\section*{Acknowledgments}
\footnotesize{Shreshth Tuli is supported by the President's PhD scholarship at Imperial College London. }

\bibliographystyle{IEEEtran}
\bibliography{references}

\begin{thebibliography}{10}
\providecommand{\url}[1]{#1}
\csname url@samestyle\endcsname
\providecommand{\newblock}{\relax}
\providecommand{\bibinfo}[2]{#2}
\providecommand{\BIBentrySTDinterwordspacing}{\spaceskip=0pt\relax}
\providecommand{\BIBentryALTinterwordstretchfactor}{4}
\providecommand{\BIBentryALTinterwordspacing}{\spaceskip=\fontdimen2\font plus
\BIBentryALTinterwordstretchfactor\fontdimen3\font minus
  \fontdimen4\font\relax}
\providecommand{\BIBforeignlanguage}[2]{{%
\expandafter\ifx\csname l@#1\endcsname\relax
\typeout{** WARNING: IEEEtran.bst: No hyphenation pattern has been}%
\typeout{** loaded for the language `#1'. Using the pattern for}%
\typeout{** the default language instead.}%
\else
\language=\csname l@#1\endcsname
\fi
#2}}
\providecommand{\BIBdecl}{\relax}
\BIBdecl

\bibitem{xu2018improved}
R.~Xu, Y.~Wang, Y.~Cheng, Y.~Zhu, Y.~Xie, A.~S. Sani, and D.~Yuan, ``Improved
  particle swarm optimization based workflow scheduling in cloud-fog
  environment,'' in \emph{International Conference on Business Process
  Management}.\hskip 1em plus 0.5em minus 0.4em\relax Springer, 2018, pp.
  337--347.

\bibitem{gill2019transformative}
S.~S. Gill, S.~Tuli, M.~Xu, I.~Singh, K.~V. Singh, D.~Lindsay, S.~Tuli,
  D.~Smirnova, M.~Singh, U.~Jain \emph{et~al.}, ``{Transformative effects of
  IoT, Blockchain and Artificial Intelligence on cloud computing: Evolution,
  vision, trends and open challenges},'' \emph{Internet of Things}, vol.~8, pp.
  100--118, 2019.

\bibitem{tuli2019fogbus}
S.~Tuli, R.~Mahmud, S.~Tuli, and R.~Buyya, ``Fogbus: A blockchain-based
  lightweight framework for edge and fog computing,'' \emph{Journal of Systems
  and Software}, vol. 154, pp. 22--36, 2019.

\bibitem{impso}
P.~Wang, Y.~Lei, P.~R. Agbedanu, and Z.~Zhang, ``Makespan-driven workflow
  scheduling in clouds using immune-based pso algorithm,'' \emph{IEEE Access},
  vol.~8, pp. 29\,281--29\,290, 2020.

\bibitem{dnsga}
G.~Ismayilov and H.~R. Topcuoglu, ``Neural network based multi-objective
  evolutionary algorithm for dynamic workflow scheduling in cloud computing,''
  \emph{Future Generation computer systems}, vol. 102, pp. 307--322, 2020.

\bibitem{adhikari2019survey}
M.~Adhikari, T.~Amgoth, and S.~N. Srirama, ``A survey on scheduling strategies
  for workflows in cloud environment and emerging trends,'' \emph{ACM Computing
  Surveys (CSUR)}, vol.~52, no.~4, pp. 1--36, 2019.

\bibitem{tuli2021cosco}
S.~Tuli, S.~R. Poojara, S.~N. Srirama, G.~Casale, and N.~R. Jennings, ``{COSCO:
  Container Orchestration Using Co-Simulation and Gradient Based Optimization
  for Fog Computing Environments},'' \emph{IEEE Transactions on Parallel and
  Distributed Systems}, vol.~33, no.~1, pp. 101--116, 2022.

\bibitem{ding2018cost}
R.~Ding, X.~Li, X.~Liu, and J.~Xu, ``A cost-effective time-constrained
  multi-workflow scheduling strategy in fog computing,'' in \emph{International
  Conference on Service-Oriented Computing}.\hskip 1em plus 0.5em minus
  0.4em\relax Springer, 2018, pp. 194--207.

\bibitem{matrouk2021scheduling}
K.~Matrouk and K.~Alatoun, ``Scheduling algorithms in fog computing: A
  survey,'' \emph{International Journal of Networked and Distributed
  Computing}, vol.~9, no.~1, pp. 59--74, 2021.

\bibitem{machen2016migrating}
A.~Machen, S.~Wang, K.~K. Leung, B.~J. Ko, and T.~Salonidis, ``Migrating
  running applications across mobile edge clouds: poster,'' in
  \emph{Proceedings of the 22nd Annual International Conference on Mobile
  Computing and Networking}, 2016, pp. 435--436.

\bibitem{ben2020edge}
A.~J. Ben~Ali, Z.~S. Hashemifar, and K.~Dantu, ``Edge-slam: Edge-assisted
  visual simultaneous localization and mapping,'' in \emph{Proceedings of the
  18th International Conference on Mobile Systems, Applications, and Services},
  2020, pp. 325--337.

\bibitem{haussmann2019cost}
J.~Haussmann, W.~Blochinger, and W.~Kuechlin, ``Cost-optimized parallel
  computations using volatile cloud resources,'' in \emph{International
  Conference on the Economics of Grids, Clouds, Systems, and Services}.\hskip
  1em plus 0.5em minus 0.4em\relax Springer, 2019, pp. 45--53.

\bibitem{choi2019optimizing}
Y.~Choi, S.~Park, and H.~Cha, ``Optimizing energy efficiency of browsers in
  energy-aware scheduling-enabled mobile devices,'' in \emph{The 25th Annual
  International Conference on Mobile Computing and Networking}, 2019, pp.
  1--16.

\bibitem{tuli2021generative}
S.~Tuli, S.~Tuli, G.~Casale, and N.~R. Jennings, ``Generative optimization
  networks for memory efficient data generation,'' \emph{Advances in Neural
  Information Processing Systems, Workshop on ML for Systems}, 2021.

\bibitem{yu2005cost}
J.~Yu, R.~Buyya, and C.~K. Tham, ``Cost-based scheduling of scientific workflow
  applications on utility grids,'' in \emph{First International Conference on
  e-Science and Grid Computing (e-Science'05)}.\hskip 1em plus 0.5em minus
  0.4em\relax Ieee, 2005, pp. 8--pp.

\bibitem{tuli2021pregan}
S.~Tuli, G.~Casale, and N.~R. Jennings, ``{PreGAN: Preemptive Migration
  Prediction Network for Proactive Fault-Tolerant Edge Computing},'' in
  \emph{IEEE Conference on Computer Communications (INFOCOM)}.\hskip 1em plus
  0.5em minus 0.4em\relax IEEE, 2022.

\bibitem{esvr}
T.-P. Pham and T.~Fahringer, ``Evolutionary multi-objective workflow scheduling
  for volatile resources in the cloud,'' \emph{IEEE Transactions on Cloud
  Computing}, 2020.

\bibitem{closure}
Y.~Wang, Y.~Guo, Z.~Guo, T.~Baker, and W.~Liu, ``Closure: A cloud scientific
  workflow scheduling algorithm based on attack--defense game model,''
  \emph{Future Generation Computer Systems}, vol. 111, pp. 460--474, 2020.

\bibitem{alkhanak2015cost}
E.~N. Alkhanak, S.~P. Lee, and S.~U.~R. Khan, ``Cost-aware challenges for
  workflow scheduling approaches in cloud computing environments: Taxonomy and
  opportunities,'' \emph{Future Generation Computer Systems}, vol.~50, pp.
  3--21, 2015.

\bibitem{genez2012workflow}
T.~A. Genez, L.~F. Bittencourt, and E.~R. Madeira, ``{Workflow scheduling for
  SaaS/PaaS cloud providers considering two SLA levels},'' in \emph{2012 IEEE
  Network Operations and Management Symposium}.\hskip 1em plus 0.5em minus
  0.4em\relax IEEE, 2012, pp. 906--912.

\bibitem{xu2009multiple}
M.~Xu, L.~Cui, H.~Wang, and Y.~Bi, ``A multiple qos constrained scheduling
  strategy of multiple workflows for cloud computing,'' in \emph{2009 IEEE
  International Symposium on Parallel and Distributed Processing with
  Applications}.\hskip 1em plus 0.5em minus 0.4em\relax IEEE, 2009, pp.
  629--634.

\bibitem{saeedi2020improved}
S.~Saeedi, R.~Khorsand, S.~G. Bidgoli, and M.~Ramezanpour, ``Improved
  many-objective particle swarm optimization algorithm for scientific workflow
  scheduling in cloud computing,'' \emph{Computers \& Industrial Engineering},
  vol. 147, p. 106649, 2020.

\bibitem{sutton2018reinforcement}
R.~S. Sutton and A.~G. Barto, \emph{Reinforcement learning: An
  introduction}.\hskip 1em plus 0.5em minus 0.4em\relax MIT press, 2018.

\bibitem{song2020win}
Z.~Song and E.~Tilevich, ``Win with what you have: Qos-consistent edge services
  with unreliable and dynamic resources,'' in \emph{2020 IEEE 40th
  International Conference on Distributed Computing Systems (ICDCS)}.\hskip 1em
  plus 0.5em minus 0.4em\relax IEEE, 2020, pp. 530--540.

\bibitem{byun2011cost}
E.-K. Byun, Y.-S. Kee, J.-S. Kim, and S.~Maeng, ``Cost optimized provisioning
  of elastic resources for application workflows,'' \emph{Future Generation
  Computer Systems}, vol.~27, no.~8, pp. 1011--1026, 2011.

\bibitem{chen2017efficient}
W.~Chen, G.~Xie, R.~Li, Y.~Bai, C.~Fan, and K.~Li, ``Efficient task scheduling
  for budget constrained parallel applications on heterogeneous cloud computing
  systems,'' \emph{Future Generation Computer Systems}, vol.~74, pp. 1--11,
  2017.

\bibitem{chirkin2017execution}
A.~M. Chirkin, A.~S. Belloum, S.~V. Kovalchuk, M.~X. Makkes, M.~A. Melnik,
  A.~A. Visheratin, and D.~A. Nasonov, ``Execution time estimation for workflow
  scheduling,'' \emph{Future generation computer systems}, vol.~75, pp.
  376--387, 2017.

\bibitem{singh2018novel}
V.~Singh, I.~Gupta, and P.~K. Jana, ``A novel cost-efficient approach for
  deadline-constrained workflow scheduling by dynamic provisioning of
  resources,'' \emph{Future Generation Computer Systems}, vol.~79, pp. 95--110,
  2018.

\bibitem{pandey2010particle}
S.~Pandey, L.~Wu, S.~M. Guru, and R.~Buyya, ``A particle swarm
  optimization-based heuristic for scheduling workflow applications in cloud
  computing environments,'' in \emph{2010 24th IEEE international conference on
  advanced information networking and applications}.\hskip 1em plus 0.5em minus
  0.4em\relax IEEE, 2010, pp. 400--407.

\bibitem{topcuoglu2002performance}
H.~Topcuoglu, S.~Hariri, and M.-Y. Wu, ``Performance-effective and
  low-complexity task scheduling for heterogeneous computing,'' \emph{IEEE
  transactions on parallel and distributed systems}, vol.~13, no.~3, pp.
  260--274, 2002.

\bibitem{ahuja2012survey}
S.~P. Ahuja, S.~Mani, and J.~Zambrano, ``A survey of the state of cloud
  computing in healthcare,'' \emph{Network and Communication Technologies},
  vol.~1, no.~2, p.~12, 2012.

\bibitem{basu2019learn}
D.~Basu, X.~Wang, Y.~Hong, H.~Chen, and S.~Bressan, ``Learn-as-you-go with
  megh: Efficient live migration of virtual machines,'' \emph{IEEE Transactions
  on Parallel and Distributed Systems}, vol.~30, no.~8, pp. 1786--1801, 2019.

\bibitem{tuli2021start}
S.~Tuli, S.~S. Gill, P.~Garraghan, R.~Buyya, G.~Casale, and N.~Jennings,
  ``{START: Straggler Prediction and Mitigation for Cloud Computing
  Environments using Encoder LSTM Networks},'' \emph{IEEE Transactions on
  Services Computing}, 2021.

\bibitem{li2015gated}
Y.~Li, D.~Tarlow, M.~Brockschmidt, and R.~Zemel, ``Gated graph sequence neural
  networks,'' \emph{arXiv preprint arXiv:1511.05493}, 2015.

\bibitem{tuli2021hunter}
S.~Tuli, S.~S. Gill, M.~Xu, P.~Garraghan, R.~Bahsoon, S.~Dustdar,
  R.~Sakellariou, O.~Rana, R.~Buyya, G.~Casale \emph{et~al.}, ``{HUNTER: AI
  based Holistic Resource Management for Sustainable Cloud Computing},''
  \emph{Journal of Systems and Software}, 2021.

\bibitem{bahdanau2015neural}
D.~Bahdanau, K.~H. Cho, and Y.~Bengio, ``Neural machine translation by jointly
  learning to align and translate,'' in \emph{3rd International Conference on
  Learning Representations, ICLR 2015}, 2015.

\bibitem{vaswani2017attention}
A.~Vaswani, N.~Shazeer, N.~Parmar, J.~Uszkoreit, L.~Jones, A.~N. Gomez,
  {\L}.~Kaiser, and I.~Polosukhin, ``Attention is all you need,'' in
  \emph{Proceedings of the 31st International Conference on Neural Information
  Processing Systems}, 2017, pp. 6000--6010.

\bibitem{lee2019attention}
J.~B. Lee, R.~A. Rossi, S.~Kim, N.~K. Ahmed, and E.~Koh, ``Attention models in
  graphs: A survey,'' \emph{ACM Transactions on Knowledge Discovery from Data
  (TKDD)}, vol.~13, no.~6, pp. 1--25, 2019.

\bibitem{huang2017arbitrary}
X.~Huang and S.~Belongie, ``Arbitrary style transfer in real-time with adaptive
  instance normalization,'' in \emph{Proceedings of the IEEE International
  Conference on Computer Vision}, 2017, pp. 1501--1510.

\bibitem{silver2009monte}
D.~Silver and G.~Tesauro, ``Monte-carlo simulation balancing,'' in
  \emph{Proceedings of the 26th Annual International Conference on Machine
  Learning}, 2009, pp. 945--952.

\bibitem{aima}
S.~Russell and P.~Norvig, \emph{Artificial Intelligence: A Modern Approach},
  3rd~ed.\hskip 1em plus 0.5em minus 0.4em\relax USA: Prentice Hall Press,
  2009.

\bibitem{rolnick2019experience}
D.~Rolnick, A.~Ahuja, J.~Schwarz, T.~P. Lillicrap, and G.~Wayne, ``Experience
  replay for continual learning,'' in \emph{Proceedings of the 33rd
  International Conference on Neural Information Processing Systems}, 2019, pp.
  350--360.

\bibitem{tuli2020dynamic}
S.~Tuli, S.~Ilager, K.~Ramamohanarao, and R.~Buyya, ``{Dynamic Scheduling for
  Stochastic Edge-Cloud Computing Environments using A3C learning and Residual
  Recurrent Neural Networks},'' \emph{IEEE Transactions on Mobile Computing},
  2020.

\bibitem{krajzewicz2012recent}
D.~Krajzewicz, J.~Erdmann, M.~Behrisch, and L.~Bieker, ``Recent development and
  applications of sumo-simulation of urban mobility,'' \emph{International
  journal on advances in systems and measurements}, vol.~5, no. 3\&4, 2012.

\bibitem{ahmed2018docker}
A.~Ahmed and G.~Pierre, ``Docker container deployment in fog computing
  infrastructures,'' in \emph{2018 IEEE International Conference on Edge
  Computing (EDGE)}.\hskip 1em plus 0.5em minus 0.4em\relax IEEE, 2018, pp.
  1--8.

\bibitem{coleman2021wfcommons}
T.~Coleman, H.~Casanova, L.~Pottier, M.~Kaushik, E.~Deelman, and R.~F.
  da~Silva, ``Wfcommons: A framework for enabling scientific workflow research
  and development,'' \emph{arXiv preprint arXiv:2105.14352}, 2021.

\bibitem{bharathi2008characterization}
S.~Bharathi, A.~Chervenak, E.~Deelman, G.~Mehta, M.-H. Su, and K.~Vahi,
  ``Characterization of scientific workflows,'' in \emph{2008 third workshop on
  workflows in support of large-scale science}.\hskip 1em plus 0.5em minus
  0.4em\relax IEEE, 2008, pp. 1--10.

\bibitem{rafique2020complementing}
W.~Rafique, L.~Qi, I.~Yaqoob, M.~Imran, R.~U. Rasool, and W.~Dou,
  ``{Complementing IoT services through software defined networking and edge
  computing: A comprehensive survey},'' \emph{IEEE Communications Surveys \&
  Tutorials}, vol.~22, no.~3, pp. 1761--1804, 2020.

\bibitem{o2019edge}
M.~O'Grady, D.~Langton, and G.~O'Hare, ``Edge computing: A tractable model for
  smart agriculture?'' \emph{Artificial Intelligence in Agriculture}, vol.~3,
  pp. 42--51, 2019.

\bibitem{semenov2021elastic}
A.~Semenov, ``Elastic computing self-organizing for artificial intelligence
  space exploration,'' in \emph{Journal of Physics: Conference Series}, vol.
  1925, no.~1.\hskip 1em plus 0.5em minus 0.4em\relax IOP Publishing, 2021, p.
  012071.

\bibitem{lin2019time}
B.~Lin, F.~Zhu, J.~Zhang, J.~Chen, X.~Chen, N.~N. Xiong, and J.~L. Mauri, ``A
  time-driven data placement strategy for a scientific workflow combining edge
  computing and cloud computing,'' \emph{IEEE Transactions on Industrial
  Informatics}, vol.~15, no.~7, pp. 4254--4265, 2019.

\bibitem{shao2019cost}
Y.~Shao, C.~Li, Z.~Fu, L.~Jia, and Y.~Luo, ``Cost-effective replication
  management and scheduling in edge computing,'' \emph{Journal of Network and
  Computer Applications}, vol. 129, pp. 46--61, 2019.

\bibitem{xie2019novel}
Y.~Xie, Y.~Zhu, Y.~Wang, Y.~Cheng, R.~Xu, A.~S. Sani, D.~Yuan, and Y.~Yang, ``A
  novel directional and non-local-convergent particle swarm optimization based
  workflow scheduling in cloud--edge environment,'' \emph{Future Generation
  Computer Systems}, vol.~97, pp. 361--378, 2019.

\bibitem{shen2015statisticalBitBrain}
S.~Shen, V.~van Beek, and A.~Iosup, ``Statistical characterization of
  business-critical workloads hosted in cloud datacenters,'' in \emph{15th
  IEEE/ACM International Symposium on Cluster, Cloud and Grid Computing}.\hskip
  1em plus 0.5em minus 0.4em\relax IEEE, 2015, pp. 465--474.

\bibitem{paszke2017automatic}
A.~Paszke, S.~Gross, S.~Chintala, G.~Chanan, E.~Yang, Z.~DeVito, Z.~Lin,
  A.~Desmaison, L.~Antiga, and A.~Lerer, ``Automatic differentiation in
  pytorch,'' 2017.

\bibitem{yates2015ensembl}
A.~Yates, K.~Beal, S.~Keenan, W.~McLaren, M.~Pignatelli, G.~R. Ritchie,
  M.~Ruffier, K.~Taylor, A.~Vullo, and P.~Flicek, ``The ensembl rest api:
  Ensembl data for any language,'' \emph{Bioinformatics}, vol.~31, no.~1, pp.
  143--145, 2015.

\bibitem{grinberg2018flask}
M.~Grinberg, \emph{Flask web development: developing web applications with
  python}.\hskip 1em plus 0.5em minus 0.4em\relax " O'Reilly Media, Inc.",
  2018.

\bibitem{adamw}
I.~Loshchilov and F.~Hutter, ``Decoupled weight decay regularization,'' in
  \emph{International Conference on Learning Representations}, 2018.

\bibitem{yi2015survey}
S.~Yi, C.~Li, and Q.~Li, ``A survey of fog computing: concepts, applications
  and issues,'' in \emph{Proceedings of the 2015 workshop on mobile big data},
  2015, pp. 37--42.

\end{thebibliography}

\begin{IEEEbiography}
[{\includegraphics[width=1in,height=1in,clip,keepaspectratio]{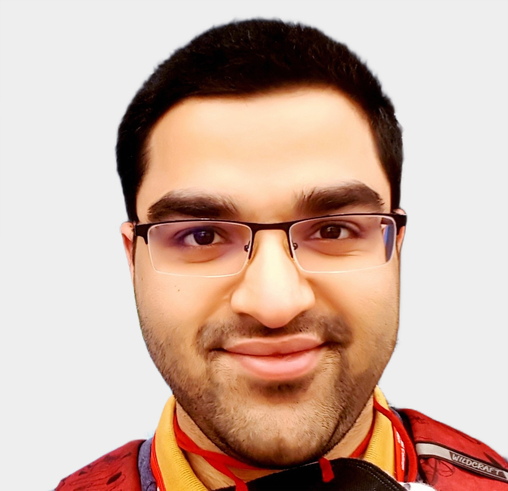}}]
{Shreshth Tuli}
is a President's Ph.D. Scholar at the Department of Computing, Imperial College London, UK. Prior to this he was an undergraduate student at the Department of Computer Science and Engineering at Indian Institute of Technology - Delhi, India. He has worked as a visiting research fellow at the CLOUDS Laboratory, School of Computing and Information Systems, the University of Melbourne, Australia. He is a national level Kishore Vaigyanik Protsahan Yojana (KVPY) scholarship holder from the Government of India for excellence in science and innovation. His research interests include Fog Computing and Deep Learning. For further information, visit \url{https://shreshthtuli.github.io/}.
\end{IEEEbiography}
\begin{IEEEbiography}
[{\includegraphics[width=1in,height=1.25in,clip,keepaspectratio]{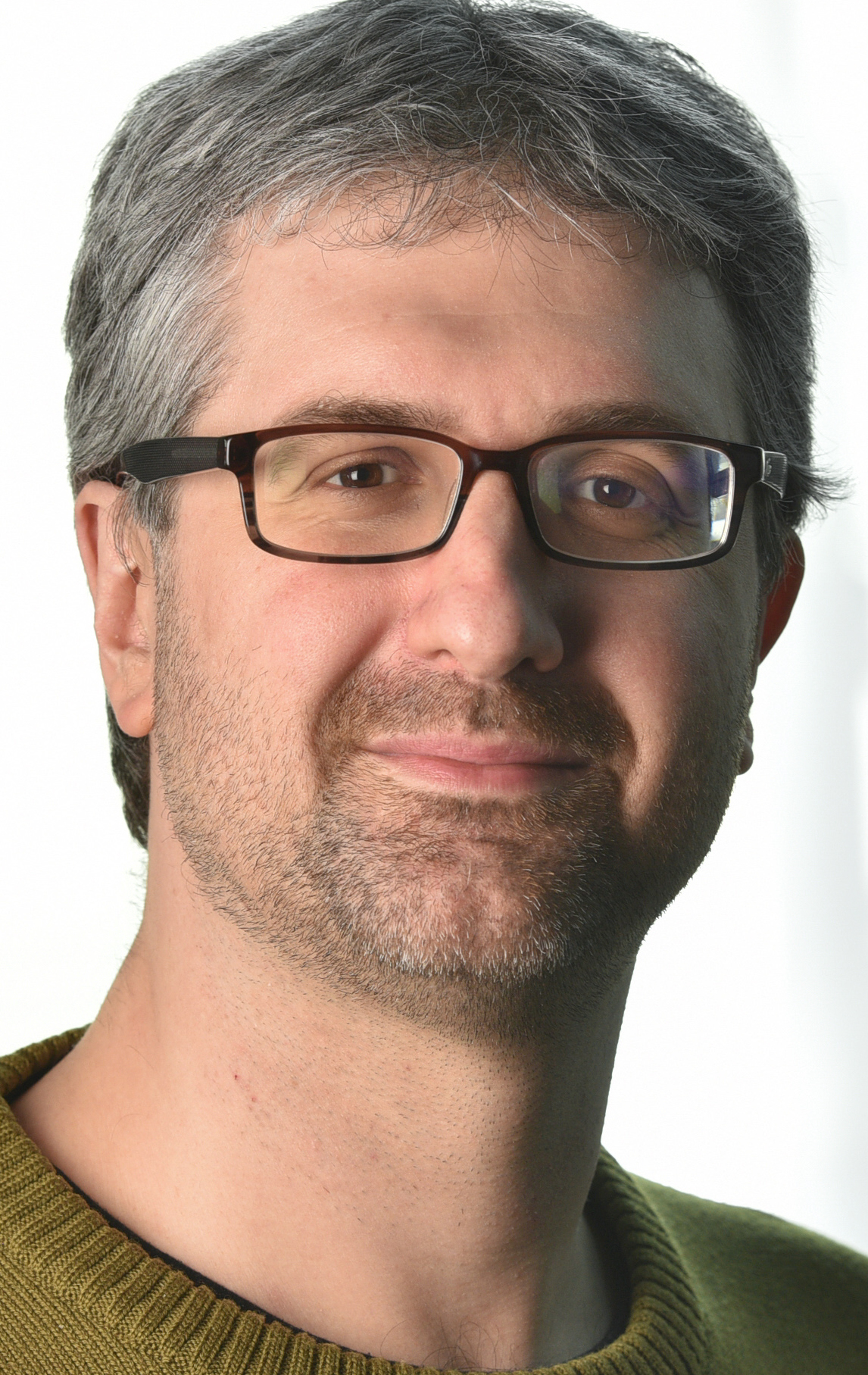}}]
{Giuliano Casale}
joined the Department of Computing at Imperial College London in 2010, where he is currently a Reader. Previously, he worked as a research scientist and consultant in the capacity planning industry. He teaches and does research in performance engineering and cloud computing, topics on which he has published more than 100 refereed papers. He has served on the technical program committee of over 80 conferences and workshops and as co-chair for several conferences in the area of performance and reliability engineering, such as ACM SIGMETRICS/Performance and IEEE/IFIP DSN. His research work has received multiple awards, recently the best paper award at ACM SIGMETRICS. He serves on the editorial boards of IEEE TNSM and ACM TOMPECS and as current chair of ACM SIGMETRICS.
\end{IEEEbiography}
\begin{IEEEbiography}
[{\includegraphics[width=1in,height=1.25in,clip,keepaspectratio]{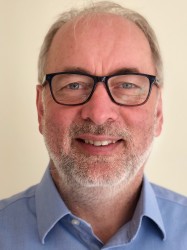}}]
{Nicholas R. Jennings}
is the Vice-Chancellor and President of Loughborough University. He is an internationally-recognised authority in the areas of AI, autonomous systems, cyber-security and agent-based computing. He is a member of the UK government’s AI Council, the governing body of the Engineering and Physical Sciences Research Council, and chair of the Royal Academy of Engineering’s Policy Committee.  Before Loughborough, he was the Vice-Provost for Research and Enterprise and Professor of Artificial Intelligence at Imperial College London, the UK's first Regius Professor of Computer Science (a post bestowed by the monarch to recognise exceptionally high quality research) and the UK Government’s first Chief Scientific Advisor for National Security.
\end{IEEEbiography}

\vfill

\end{document}